\documentclass[longauth]{aa}
\usepackage{graphicx}
\usepackage{txfonts}
\usepackage{xcolor}
\usepackage{hyperref}
\hypersetup{colorlinks, linkcolor=blue, citecolor=blue, urlcolor=blue}
\usepackage{amsmath}
\usepackage{ulem}
\usepackage{soul}
\usepackage{enumitem}

\newcommand{\te}{t_{\rm E}}
\newcommand{\thetae}{\theta_{\rm E}}

\newcommand{\pie}{\pi_{\rm E}}

\newcommand{\dl}{D_{\rm L}}
\newcommand{\ds}{D_{\rm S}}

\definecolor{brown}{rgb}{0.59, 0.29, 0.0}
\definecolor{darkgreen}{rgb}{0.0, 0.42, 0.24}
\definecolor{darkblue}{rgb}{0.01, 0.31, 0.59}
\definecolor{darkblue}{rgb}{0.0, 0.25, 0.42}
\definecolor{blue}{rgb}{0.0,0.0,1.0}
\definecolor{green}{rgb}{0.0,1.0,0.0}

\begin{document}

\title{Microlensing brown-dwarf companions in binaries detected during the 2022 and 2023 seasons}
\titlerunning{Microlensing brown dwarfs during the 2022 and 2023 seasons}

\author{
     Cheongho~Han\inst{\ref{inst1}} 
\and Ian~A.~Bond\inst{\ref{inst2}}
\and Andrzej~Udalski\inst{\ref{inst3}} 
\and Chung-Uk~Lee\inst{\ref{inst4},\ref{inst25}} 
\and Andrew~Gould\inst{\ref{inst5},\ref{inst6}}      
\\
(Leading authors)
\\
     Michael~D.~Albrow\inst{\ref{inst7}}   
\and Sun-Ju~Chung\inst{\ref{inst4}}      
\and Kyu-Ha~Hwang\inst{\ref{inst4}} 
\and Youn~Kil~Jung\inst{\ref{inst4}} 
\and Yoon-Hyun~Ryu\inst{\ref{inst4}} 
\and Yossi~Shvartzvald\inst{\ref{inst8}}   
\and In-Gu~Shin\inst{\ref{inst9}} 
\and Jennifer~C.~Yee\inst{\ref{inst9}}   
\and Hongjing~Yang\inst{\ref{inst10}}     
\and Weicheng~Zang\inst{\ref{inst9},\ref{inst10}}     
\and Sang-Mok~Cha\inst{\ref{inst4},\ref{inst11}} 
\and Doeon~Kim\inst{\ref{inst1}}
\and Dong-Jin~Kim\inst{\ref{inst4}} 
\and Seung-Lee~Kim\inst{\ref{inst4}} 
\and Dong-Joo~Lee\inst{\ref{inst4}} 
\and Yongseok~Lee\inst{\ref{inst4},\ref{inst11}} 
\and Byeong-Gon~Park\inst{\ref{inst4}} 
\and Richard~W.~Pogge\inst{\ref{inst6}}
\\
(The KMTNet Collaboration)
\\
     Fumio~Abe\inst{\ref{inst12}}
\and Ken~Bando\inst{\ref{inst13}}
\and Richard~Barry\inst{\ref{inst14}}
\and David~P.~Bennett\inst{\ref{inst14},\ref{inst15}}
\and Aparna~Bhattacharya\inst{\ref{inst14},\ref{inst15}}
\and Hirosame~Fujii\inst{\ref{inst12}}
\and Akihiko~Fukui\inst{\ref{inst16},}\inst{\ref{inst17}}
\and Ryusei~Hamada\inst{\ref{inst13}}
\and Shunya~Hamada\inst{\ref{inst13}}
\and Naoto~Hamasaki\inst{\ref{inst13}}
\and Yuki~Hirao\inst{\ref{inst18}}
\and Stela~Ishitani Silva\inst{\ref{inst14},\ref{inst19}}
\and Yoshitaka~Itow\inst{\ref{inst12}}
\and Rintaro~Kirikawa\inst{\ref{inst13}}
\and Naoki~Koshimoto\inst{\ref{inst13}}
\and Yutaka~Matsubara\inst{\ref{inst12}}
\and Shota~Miyazaki\inst{\ref{inst20}}
\and Yasushi~Muraki\inst{\ref{inst12}}
\and Tutumi~Nagai\inst{\ref{inst13}}
\and Kansuke~Nunota\inst{\ref{inst13}}
\and Greg~Olmschenk\inst{\ref{inst14}}
\and Cl{\'e}ment~Ranc\inst{\ref{inst21}}
\and Nicholas~J.~Rattenbury\inst{\ref{inst22}}
\and Yuki~Satoh\inst{\ref{inst13}}
\and Takahiro~Sumi\inst{\ref{inst13}}
\and Daisuke~Suzuki\inst{\ref{inst13}}
\and Mio~Tomoyoshi\inst{\ref{inst13}}
\and Paul~J.~Tristram\inst{\ref{inst23}}
\and Aikaterini~Vandorou\inst{\ref{inst14},\ref{inst15}}
\and Hibiki~Yama\inst{\ref{inst13}}
\and Kansuke~Yamashita\inst{\ref{inst13}}
\\
(The MOA Collaboration)
\\
     Przemek~Mr{\'o}z\inst{\ref{inst3}} 
\and Micha{\l}~K.~Szyma{\'n}ski\inst{\ref{inst3}}
\and Jan~Skowron\inst{\ref{inst3}}
\and Rados{\l}aw~Poleski\inst{\ref{inst3}} 
\and Igor~Soszy{\'n}ski\inst{\ref{inst3}}
\and Pawe{\l}~Pietrukowicz\inst{\ref{inst3}}
\and Szymon~Koz{\l}owski\inst{\ref{inst3}} 
\and Krzysztof~A.~Rybicki\inst{\ref{inst3},\ref{inst8}}
\and Patryk~Iwanek\inst{\ref{inst3}}
\and Krzysztof~Ulaczyk\inst{\ref{inst24}}
\and Marcin~Wrona\inst{\ref{inst3},\ref{inst25}}
\and Mariusz~Gromadzki\inst{\ref{inst3}}          
\and Mateusz~J.~Mr{\'o}z\inst{\ref{inst3}} 
\\
(The OGLE Collaboration)
}

\institute{
      Department of Physics, Chungbuk National University, Cheongju 28644, Republic of Korea\label{inst1}                                                          
\and  Institute of Natural and Mathematical Science, Massey University, Auckland 0745, New Zealand\label{inst2}                                                    
\and  Astronomical Observatory, University of Warsaw, Al.~Ujazdowskie 4, 00-478 Warszawa, Poland\label{inst3}                                                      
\and  Korea Astronomy and Space Science Institute, Daejon 34055, Republic of Korea\label{inst4}                                                                    
\and  Max Planck Institute for Astronomy, K\"onigstuhl 17, D-69117 Heidelberg, Germany\label{inst5}                                                                
\and  Department of Astronomy, The Ohio State University, 140 W. 18th Ave., Columbus, OH 43210, USA\label{inst6}                                                   
\and  University of Canterbury, Department of Physics and Astronomy, Private Bag 4800, Christchurch 8020, New Zealand\label{inst7}                                 
\and  Department of Particle Physics and Astrophysics, Weizmann Institute of Science, Rehovot 76100, Israel\label{inst8}                                           
\and  Center for Astrophysics $|$ Harvard \& Smithsonian 60 Garden St., Cambridge, MA 02138, USA\label{inst9}                                                      
\and  Department of Astronomy and Tsinghua Centre for Astrophysics, Tsinghua University, Beijing 100084, China\label{inst10}                                       
\and  School of Space Research, Kyung Hee University, Yongin, Kyeonggi 17104, Republic of Korea\label{inst11}                                                      
\and  Institute for Space-Earth Environmental Research, Nagoya University, Nagoya 464-8601, Japan\label{inst12}                                                    
\and  Department of Earth and Space Science, Graduate School of Science, Osaka University, Toyonaka, Osaka 560-0043, Japan\label{inst13}                           
\and  Code 667, NASA Goddard Space Flight Center, Greenbelt, MD 20771, USA\label{inst14}                                                                           
\and  Department of Astronomy, University of Maryland, College Park, MD 20742, USA\label{inst15}                                                                   
\and  Department of Earth and Planetary Science, Graduate School of Science, The University of Tokyo, 7-3-1 Hongo, Bunkyo-ku, Tokyo 113-0033, Japan\label{inst16}  
\and  Instituto de Astrof{\'i}sica de Canarias, V{\'i}a L{\'a}ctea s/n, E-38205 La Laguna, Tenerife, Spain\label{inst17}                                           
\and  Institute of Astronomy, Graduate School of Science, The University of Tokyo, 2-21-1 Osawa, Mitaka, Tokyo 181-0015, Japan\label{inst18}                       
\and  Oak Ridge Associated Universities, Oak Ridge, TN 37830, USA\label{inst19}                                                                                    
\and  Institute of Space and Astronautical Science, Japan Aerospace Exploration Agency, 3-1-1 Yoshinodai, Chuo, Sagamihara, Kanagawa 252-5210, Japan\label{inst20} 
\and  Sorbonne Universit\'e, CNRS, UMR 7095, Institut d'Astrophysique de Paris, 98 bis bd Arago, 75014 Paris, France\label{inst21}                                 
\and  Department of Physics, University of Auckland, Private Bag 92019, Auckland, New Zealand\label{inst22}                                                        
\and  University of Canterbury Mt.~John Observatory, P.O. Box 56, Lake Tekapo 8770, New Zealand\label{inst23}                                                      
\and  Department of Physics, University of Warwick, Gibbet Hill Road, Coventry, CV4 7AL, UK\label{inst24}                                                          
\and  Villanova University, Department of Astrophysics and Planetary Sciences, 800 Lancaster Ave., Villanova, PA 19085, USA                                        
\and  Corresponding author\label{inst25}                                                                                                                           
}                                                                                                                                                       
\date{Received ; accepted}

\abstract
{}
{
Building on previous works to construct a homogeneous sample of brown dwarfs in binary
systems, we investigate microlensing events detected by the Korea Microlensing Telescope Network
(KMTNet) survey during the 2022 and 2023 seasons.
}
{
Given the difficulty in distinguishing brown-dwarf events from those produced by binary lenses
with nearly equal-mass components, we analyze all lensing events detected during the seasons that
exhibit anomalies characteristic of binary-lens systems.	
}
{
Using the same criteria consistently applied in previous studies, we identify six additional 
brown dwarf candidates through the analysis of lensing events KMT-2022-BLG-0412, KMT-2022-BLG-2286, 
KMT-2023-BLG-0201, KMT-2023-BLG-0601, KMT-2023-BLG-1684, and KMT-2023-BLG-1743. An examination of 
the mass posteriors shows that the median mass of the lens companions ranges from 0.02\,$M_\odot$ 
to 0.05\,$M_\odot$, indicating that these companions fall within the brown-dwarf mass range. The 
mass of the primary lenses ranges from 0.11\,$M_\odot$ to 0.68\,$M_\odot$, indicating that they 
are low-mass stars with substantially lower masses compared to the Sun.
}
{}


\maketitle

\section{Introduction}\label{sec:one}

Brown dwarfs (BDs) are substellar objects capable of undergoing deuterium fusion, yet lacking 
the mass required for sustaining hydrogen fusion. This places them in a mass range between 
stars and planets \citep{Burrows1997, Chabrier2000}.  Understanding the binary properties of BDs, 
such as binary frequency, separation, and mass-ratio distributions, is essential for comprehending 
these objects as these properties provide insights into both their formation processes and the 
dynamic interactions within systems.  For instance, the scarcity of BDs in close orbits relative 
to the frequency of either less massive planetary companions or more massive stellar companions, 
known as the BD desert \citep{Marcy2000, Grether2006}, has been observed around solar-mass stars. 
This may suggest the possibility of inward migration of BD companions or a low probability of BD 
formation due to reduced gas accretion rates during runaway accretion. However, the existence of 
this BD desert has not been firmly established for binaries with M dwarfs, which are the most 
abundant stellar population in the Galaxy, because of the limited sample size of BD binaries. 
Therefore, expanding the sample size of binary BD systems is crucial for gaining deeper insights 
into the intricacies of BD formation.

Microlensing, as it does not rely on the luminosity of lens objects, serves as an important 
tool for detecting binaries containing faint BD companions. To create a BD binary sample 
independent of the stellar types of primary stars, we extensively have examined microlensing 
data collected by high-cadence surveys since 2016. From this investigation, we reported six 
BD binaries detected during the 2016--2018 seasons \citep[Paper I]{Han2022}, four binaries 
during the 2018--2020 seasons \citep[Paper II]{Han2023a}, and another four binaries in the 
2021 season \citep[Paper III]{Han2023b}.  The key parameter distinguishing a BD from a low-mass 
star at the upper mass bound and from large-mass planets at the lower mass bound is the mass.  
However, it is generally very difficult to determine the mass of a lens.  Therefore, in 
selecting the BD sample, the previous three studies employed a method in which the mass ratio 
between a BD companion and its stellar host is less than $q_{\rm max} = 0.1$. This approach 
was chosen because the mass ratio between the lens components can be accurately measured for 
most binary-lens single-source (2L1S) lensing events, and companions with mass ratios less than 
$q_{\rm max}$ are highly likely to be BDs, considering that most lensing events are produced 
by low-mass stars \citep{Han2003}.

Apart from the BDs documented in Papers I through III since 2016, specific BD detections were 
singled out for their own scientific significance.  In the single-lens single-source (1L1S) 
lensing event OGLE-2017-BLG-0896, the lens was identified as an isolated BD based on the measured 
lens mass, determined through microlens-parallax analysis with additional data from the space-based 
{\it Spitzer} telescope \citep{Shvartzvald2019}. Extra {\it Spitzer} observations also revealed that 
the lenses of the events MOA-2016-BLG-231 \citep{Chung2019} and OGLE-2019-BLG-0033 \citep{Herald2022} 
were binaries consisting of BD companions and M-dwarf primaries. Additionally, the lens of 
OGLE-2016-BLG-1266 was found to be a binary composed of a BD and a super-Jupiter \citep{Albrow2018}.  
For OGLE-2017-BLG-1522, the lens was identified as a planetary system in which a BD primary hosts 
a giant planet companion \citep{Jung2018}. Similarly, the lens of MOA-2015-BLG-337 was found to be 
a planetary system with a host near the BD/planetary-mass boundary or a BD binary \citep{Miyazaki2018}. 
\citet{Gould2022} reported an isolated BD from the analysis of the short timescale 1L1S event 
KMT-2022-BLG-2397. Additionally, \citet{Han2020} reported that the lenses of the short timescale 
1L1S events MOA-2017-BLG-147 and MOA-2017-BLG-241 are likely to be isolated BDs, while the lens of 
MOA-2019-BLG-256 is a binary system with both companions likely being BDs.  Another microlensing 
binary composed of two BDs was identified by determining the lens mass from the measured angular 
Einstein radius and microlens parallax of the 2L1S event OGLE-2016-BLG-1469L \citep{Han2017a}. By 
measuring the ground-based microlens parallax, it was reported that the lenses of MOA-2019-BLG-008 
\citep{Bachelet2022}, OGLE-2016-BLG-0693 \citep{Ryu2017}, and OGLE-2014-BLG-1112 \citep{Han2017b} 
are binaries composed of a main-sequence dwarf star and a BD companion. The list of microlensing 
BDs in binaries detected before the 2016 season is detailed in Table 1 of \citet{Chung2019}.

Building on the work presented in Papers I through III, we report six additional BDs in 
binary systems identified through a systematic inspection of microlensing events detected 
by the Korea Microlensing Telescope Network \citep[KMTNet;][]{Kim2016} survey during the 2022 
and 2023 seasons. To create a homogeneous sample of BD companions for future statistical 
analyses of BD properties, we applied the same criteria used in the previous detections.

\begin{table*}[t]
\caption{Source coordinates, baseline magnitude, and $I$-bband extinction.\label{table:one}}
\begin{tabular}{lllllll}
\hline\hline
\multicolumn{1}{c}{Parameter}                                    &
\multicolumn{1}{c}{$({\rm RA}, {\rm DEC})_{\rm J2000}$ }         &
\multicolumn{1}{c}{$(l, b)$}                                     &
\multicolumn{1}{c}{$I_{\rm base}$}                               &
\multicolumn{1}{c}{$A_I$}                                        \\
\hline
 KMT-2022-BLG-0412    &  (18:16:43.38, -26:36:00.29)   &   (5$^\circ\hskip-2pt$.439, -4$^\circ\hskip-2pt$.824)    &   18.63    &  0.78   \\ 
 KMT-2022-BLG-2286    &  (17:56:39.66, -28:51:17.71)   &   (1$^\circ\hskip-2pt$.305, -2$^\circ\hskip-2pt$.035)    &   19.90    &  2.10   \\ 
 KMT-2023-BLG-0201    &  (17:59:48.01, -30:29:59.89)   &   (0$^\circ\hskip-2pt$.217, -3$^\circ\hskip-2pt$.446)    &   15.54    &  1.36   \\ 
 KMT-2023-BLG-0601    &  (17:27:03.64, -29:28:13.30)   &  (-2$^\circ\hskip-2pt$.631,  3$^\circ\hskip-2pt$.126)    &   20.05    &  1.86   \\ 
 KMT-2023-BLG-1684    &  (17:35:25.77, -35:16:52.68)   &  (-6$^\circ\hskip-2pt$.523, -1$^\circ\hskip-2pt$.537)    &   20.38    &  5.41   \\ 
 KMT-2023-BLG-1743    &  (17:54:53.73, -33:10:50.99)   &  (-2$^\circ\hskip-2pt$.632, -3$^\circ\hskip-2pt$.879)    &   17.89    &  1.40   \\ 
\hline                                                                                                                 
\end{tabular}
\end{table*}

\section{Procedure of event selection}\label{sec:two}

Identifying BD companions in gravitational lensing events requires systematic analysis of 
anomalous lensing events. In a lensing event produced by a binary composed of a planet and 
its stellar host, the mass ratio between the lens components is typically less than $10^{-2}$. 
In such cases, the planetary nature of the lens can be readily identified because the anomaly 
usually appears as a short-term perturbation on the smooth 1L1S lensing light curve of the 
primary lens \citep{Mao1991, Gould1992b}.  In contrast, for 2L1S events with BD lens companions 
with mass ratios in the range of $10^{-2}\lesssim q \lesssim 10^{-1}$, anomalies can often be 
difficult to distinguish from those of 2L1S events with roughly equal lens mass components. 
Therefore, we analyze all lensing events exhibiting anomalies characteristic of 2L1S events, 
such as spikes resulting from the source crossing the lensing caustic formed by a binary lens, 
or bumps resulting from the source approaching the cusp of the caustic. Caustics are positions 
on the source plane where the lensing magnification of a point source becomes infinite. They 
form when the lens is composed of multiple objects, and thus caustic-related features in the 
lensing light curve indicate that the lens is likely to be binary.  The total number of 
microlensing events detected from the KMTNet survey was 2803 in the 2022 season and 3162 in 
the 2023 season.  Approximately 10\% of these events display anomalies of various origins, 
with about half of those anomalous events showing clear deviations involving caustics.

Modeling the light curves of 2L1S events involves searching for lensing parameters that best 
fit the observed light curves. Describing a 2L1S light curve requires seven basic parameters. 
The first three parameters $(t_0, u_0, \te)$ describe the lens-source approach, representing 
the time of the closest approach, the lens-source separation at that time (impact parameter), 
and the event timescale, respectively. The value of the impact parameter is scaled to the 
angular Einstein radius ($\thetae$) of the lens system, and the timescale is defined as the 
duration for the source to transit $\thetae$. Two additional parameters $(s, q)$ define the 
binary lens: $s$ denotes the projected separation (normalized to $\thetae$), and $q$ represents 
the mass ratio between the lens components. The sixth parameter, $\alpha$, denotes the source 
trajectory angle, defined as the angle between the source trajectory and the binary lens axis. 
The final parameter, $\rho$, is the ratio of the angular source radius $\theta_*$ to the Einstein 
radius ($\rho=\theta_*/\thetae$; normalized source radius), characterizing the deformation of 
the lensing light curve due to finite-source effects during caustic crossings and approaches.

The modeling was carried out according to the following procedure. First, we conducted searches
for the binary parameters $s$ and $q$ using a grid approach, covering a wide range that encompasses 
various populations of lens companions, including planets, BDs, and stars.  The grid search was 
conducted over the ranges $-1 \leq \log s < 1$ for the binary separation and $-5 \leq \log q < 1$ 
for the mass ratio.  During this process, modeling was done with multiple starting $\alpha$ values, 
which are divided into 21 intervals over the range $0 \leq \alpha < 2\pi$.  The remaining lensing 
parameters were optimized using a downhill method based on the Markov chain Monte Carlo (MCMC) 
algorithm, employing an adaptive step-size Gaussian sampler \citep{Doran2004}.  In the second step, 
we inspected the $\chi^2$ map in the plane of grid parameters and identified local minima. In the 
final step, we refined the individual local solutions by allowing all parameters to vary freely. 
We compared the $\chi^2$ values of the individual local solutions and presented a global solution.  
If multiple local solutions exist and there is a significant degeneracy among them, we present 
all degenerate solutions and examine the underlying reasons for this degeneracy.

For events with long timescales and well-covered light curves, we additionally conducted modeling 
considering higher-order effects.  We consider two such effects: microlens-parallax and lens-orbital 
effects. The microlens-parallax effect is caused by the Earth's orbital motion, which induces 
deviations in the relative lens-source motion from a rectilinear path \citep{Gould1992a}. The lens 
orbital motion effect arises from the orbital motion of the lens itself, which also causes deviations 
from the rectilinear relative motion between the lens and the source \citep{Albrow2000}.  Measuring 
the microlens parallax ($\pie$) by incorporating higher-order effects in the modeling is important, 
because it can further constrain the physical lens parameters of the lens mass and distance.\footnote{
None of the six events analyzed in this work showed detectable higher-order effects.  Four of these 
events had timescales shorter than a month, which is insufficient to produce noticeable deviations.  
In one of the remaining events with a relatively long timescale, the photometric precision was 
insufficient to measure the subtle deviations caused by higher-order effects.  In the other event, 
higher-order effects went undetected because the light curve was only partially observed.  } For 
computing finite-source magnifications, we employed the map-making method detailed in \citet{Dong2006}.  
In this procedure, we took into account the limb-darkening effect arising from the surface brightness 
variation of the source star.

Among the analyzed 2L1S events, we selected candidate BD 2L1S events by applying the criterion
that the mass ratio between the binary lens components lies in the range of $10^{-2} < q < 10^{-1}$.  
For short timescale events with $\te \lesssim 10$ days, we set the upper limit of the mass ratio at 
$q_{\rm max} = 0.2$. This higher threshold accounts for the fact that lens mass is proportional to 
the square of the timescale, making lower-mass lenses more likely to produce short timescale events.
We note that this criterion has been consistently applied in Papers I through III for finding BD 
companions. Therefore, the BDs identified in this and previous works form a homogeneous sample, 
which is crucial for the statistical analysis of BD properties.  From this selection criterion, we 
identified six candidate events including KMT-2022-BLG-0412, KMT-2022-BLG-2286, KMT-2023-BLG-0201, 
KMT-2023-BLG-0601, KMT-2023-BLG-1684, and KMT-2023-BLG-1743.  These candidate BD events account for 
roughly 3\% of the approximately 200 anomalous lensing events from the 2022 and 2023 seasons with 
analyzed light curves. However, this fraction does not represent the true proportion of binary lenses 
containing a BD, as the selection process for BD candidates is not automated and thus subject to human bias.

Table~\ref{table:one} presents the coordinates, baseline magnitude of source stars, and extinction 
values toward the field of the events. For the identified events, we checked whether they were 
also observed by other lensing surveys, specifically the Microlensing Observations in Astrophysics 
\citep[MOA;][]{Bond2001} and the Optical Gravitational Lensing Experiment \citep[OGLE;][]{Udalski2015}. 
We found that the OGLE group observed KMT-2023-BLG-0201, while the MOA group observed KMT-2023-BLG-0201 
and KMT-2023-BLG-1743. For these additionally observed events, we included the data in our analyses.

\section{Observations and data}\label{sec:three}

The KMTNet survey has been fully operational since 2016, following a one-year commissioning run
in the 2015 season.  The primary aim of the survey is to discover extrasolar planets through the 
observation of lensing events in stars positioned within the Galactic bulge field. The survey 
utilizes three identical telescopes specifically designed for high-cadence monitoring, aiming 
to detect planets with masses below that of Earth. To ensure continuous monitoring of lensing 
events, the KMTNet telescopes are strategically distributed across three countries in the 
Southern Hemisphere: the Cerro Tololo Inter-American Observatory in Chile (KMTC), the South 
African Astronomical Observatory in South Africa (KMTS), and the Siding Spring Observatory in 
Australia (KMTA). Each telescope features a 1.6-meter aperture and is equipped with a camera 
that offers a field of view spanning 4 square degrees. KMTNet observations primarily utilize 
the $I$ passband, with approximately one-tenth of the images taken in the $V$ band to measure 
the color of the source.

The MOA survey utilizes the 1.8-meter telescope at the Mt. John Observatory in New Zealand,
while the OGLE survey utilizes the 1.3-meter telescope at the Las Campanas Observatory in Chile.
These telescopes are equipped with cameras that provide fields of view of 1.4 and 2.2 square
degrees, respectively. While OGLE observations were conducted in the same $I$ band used by the
KMTNet survey, observations by the MOA group were conducted in a customized MOA-$R$ band.

Images from the surveys were reduced and photometry of the events was performed using the
automated pipelines specific to each survey. The pipeline employed by the KMTNet survey was
developed by \citet{Albrow2009}, while those utilized by the OGLE and MOA groups were
developed by \citet{Udalski2003} and \citet{Bond2001}, respectively. To utilize optimal data, we
performed additional photometry on the KMTNet data using the code developed by \citet{Yang2024}. 
During the modeling process, we adjusted the error bars of the data not only to normalize
the $\chi^2$ value for each data set to unity but also to ensure consistency between the error
bars and the scatter of the data. This normalization process was done following the routine
detailed in \citet{Yee2012}.

\begin{figure}[t]
\includegraphics[width=\columnwidth]{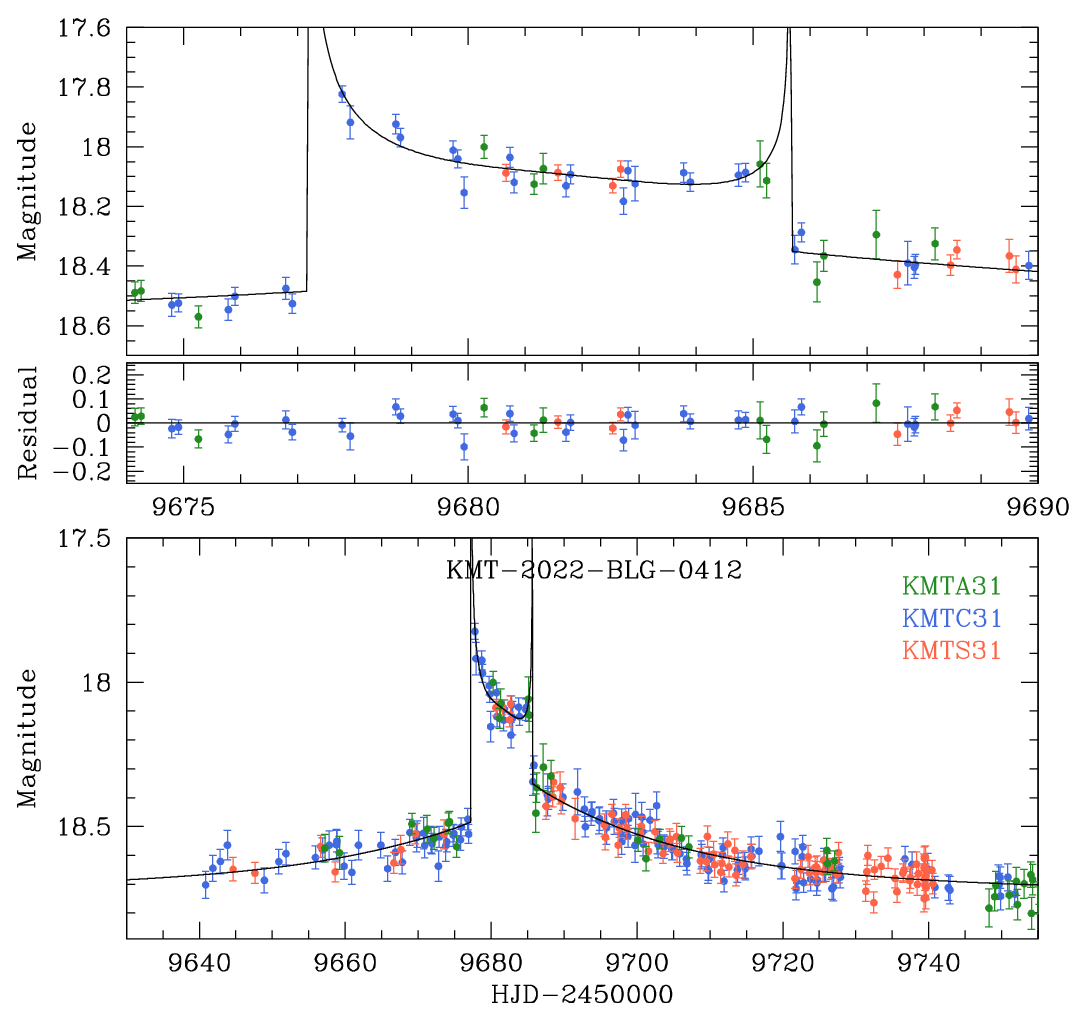}
\caption{
Lensing light curve of KMT-2022-BLG-0412.  The lower panel provides a comprehensive view, while
the upper two panels offer magnified views around the caustic-crossing features and the residuals 
from the model. The curve overlaid on the data points is the best-fit model. The colors of the 
data points are selected to correspond to the telescope labels in the legend.	
}
\label{fig:one}
\end{figure}

\section{Analyses of events}\label{sec:four}

In this section, we present the analyses performed for the individual lensing events.  For each 
event, we begin by describing the anomaly observed in the light curve, proceed to present the 
best-fit parameters, and conclude with a description of the lens system's configuration.  In 
describing the light curve, we use an abridged Heliocentric Julian Date, defined as 
${\rm HJD}^\prime \equiv {\rm HJD} - 2450000$.

\subsection{KMT-2022-BLG-0412}\label{sec:four-one}

The lensing magnification of the source flux for the lensing event KMT-2022-BLG-0412 commenced 
prior to the onset of the 2022 bulge season on March 1, 2022. The gravitational lensing nature 
of the source flux variation was discerned on April 13 (${\rm HJD}^\prime =9683$). This event 
was exclusively observed by the KMTNet group, with the source situated in the KMTNet BLG31 field, 
toward which observations were conducted at a cadence of 2.5 hours.

Figure~\ref{fig:one} displays the light curve of the event. The light curve features two spikes, 
occurring at approximately ${\rm HJD}^\prime=9677.2$ and 9685.7, caused by caustic crossings of 
the source star. Because a caustic forms a closed curve, caustic spikes appear in pairs, with 
each pair arising when the source enters and exits the caustic. Between the caustic spikes, the 
light exhibits a skewed variation resembling a "U" shape. The left and right sides of the caustic 
feature are asymmetric, with the region just after the caustic exit being approximately 0.11 
magnitude brighter than the region before the caustic entrance.

\begin{figure}[t]
\includegraphics[width=\columnwidth]{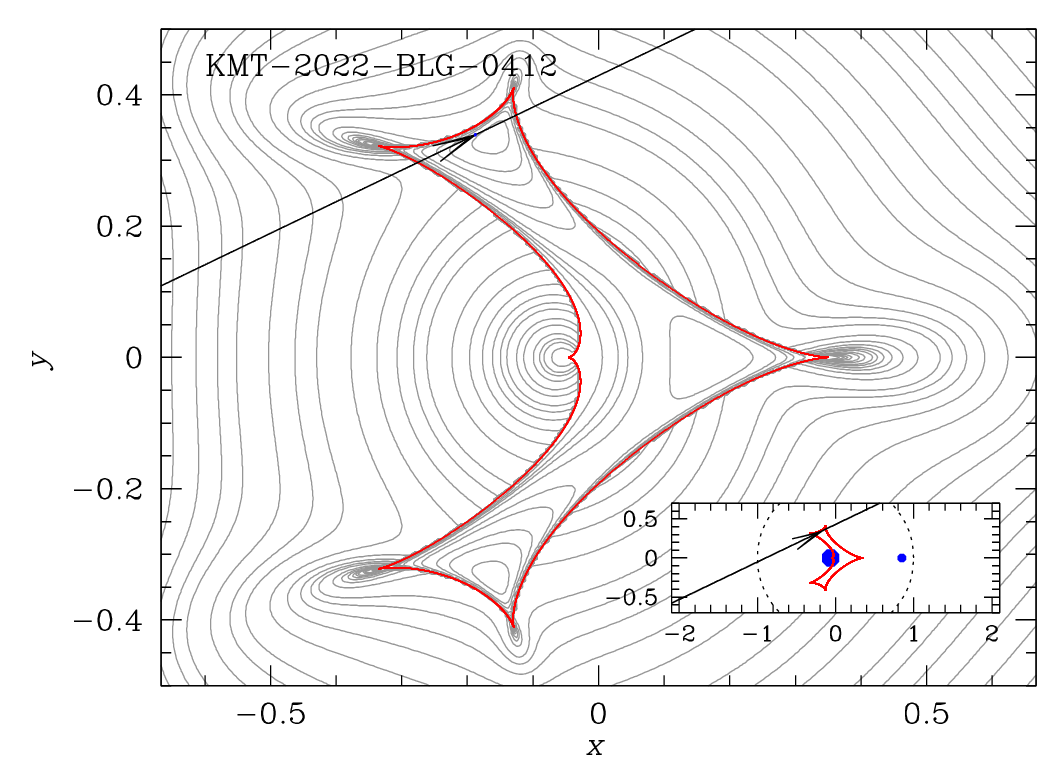}
\caption{
Configuration of the lens system for the lensing event KMT-2022-BLG-0412. The red figure
composed of concave curves represents the caustic, while the arrowed line represents the 
trajectory of the source. The grey curves encompassing the caustic represent equi-magnification 
contours.  Lengths are scaled to the angular Einstein radius of the lens system. The inset shows 
the comprehensive view including the positions of the lens components (marked by blue dots) and 
Einstein radius (dotted circle).
}
\label{fig:two}
\end{figure}

\begin{table}[t]
\caption{Lensing parameters of KMT-2022-BLG-0412.\label{table:two}}
\begin{tabular*}{\columnwidth}{@{\extracolsep{\fill}}llll}
\hline\hline
\multicolumn{1}{c}{Parameter}   &
\multicolumn{1}{c}{Value}        \\
\hline
$\chi^2$                &  $496.5             $   \\
$t_0$ (HJD$^\prime$)    &  $9683.289 \pm 1.002$   \\
$u_0$                   &  $0.387 \pm 0.030   $   \\
$\te$ (days)            &  $50.75 \pm 2.14    $   \\
$s$                     &  $0.9124 \pm 0.0091 $   \\
$q$                     &  $0.075 \pm 0.017   $   \\
$\alpha$ (rad)          &  $2.694 \pm 0.076   $   \\
$\rho$ ($10^{-3}$)      &  $ < 4              $   \\
\hline
\end{tabular*}
\end{table}

From the modeling the observed light curve, we identified a unique solution. In Table~\ref{table:two}, 
we present the lensing parameters of the solution along with the $\chi^2$ value of the model fit. 
The model curve is depicted in Figure~\ref{fig:one}. The lower panel offers a comprehensive view, 
while the upper two panels provide magnified views of the caustic-crossing features and the residuals 
from the model.  The limited coverage of both caustic spikes hindered the precise determination of 
the normalized source radius, resulting in only an upper limit being established.  The parameters 
characterizing the binary lens are $(s, q)\sim (0.91, 0.075)$, indicating the presence of a low-mass 
companion with a projected separation slightly less than the Einstein radius. The timescale of the 
event, $\te \sim 51$~days, is fairly long, constituting a sizable portion of the Earth's orbital 
period. However, determining the microlens parallax proved difficult, primarily due to substantial 
photometric uncertainties arising from the faintness of the source.

Figure~\ref{fig:two} illustrates the lens-system configuration, showing the source's trajectory 
with respect to the caustic formed by the binary lens. The companion is positioned slightly inside 
the Einstein ring. Despite the low mass ratio, the lens companion generates a singular, expansive 
resonant caustic due to the binary separation's close proximity to unity. This caustic consists of 
six folds, with the two caustic spikes occurring as the source traversed the upper two folds that 
connect the central and outer regions of the caustic. The asymptotic approach of the source near 
the upper fold led to the skewed flux variation observed in the region between the caustic spikes.

\begin{figure}[t]
\includegraphics[width=\columnwidth]{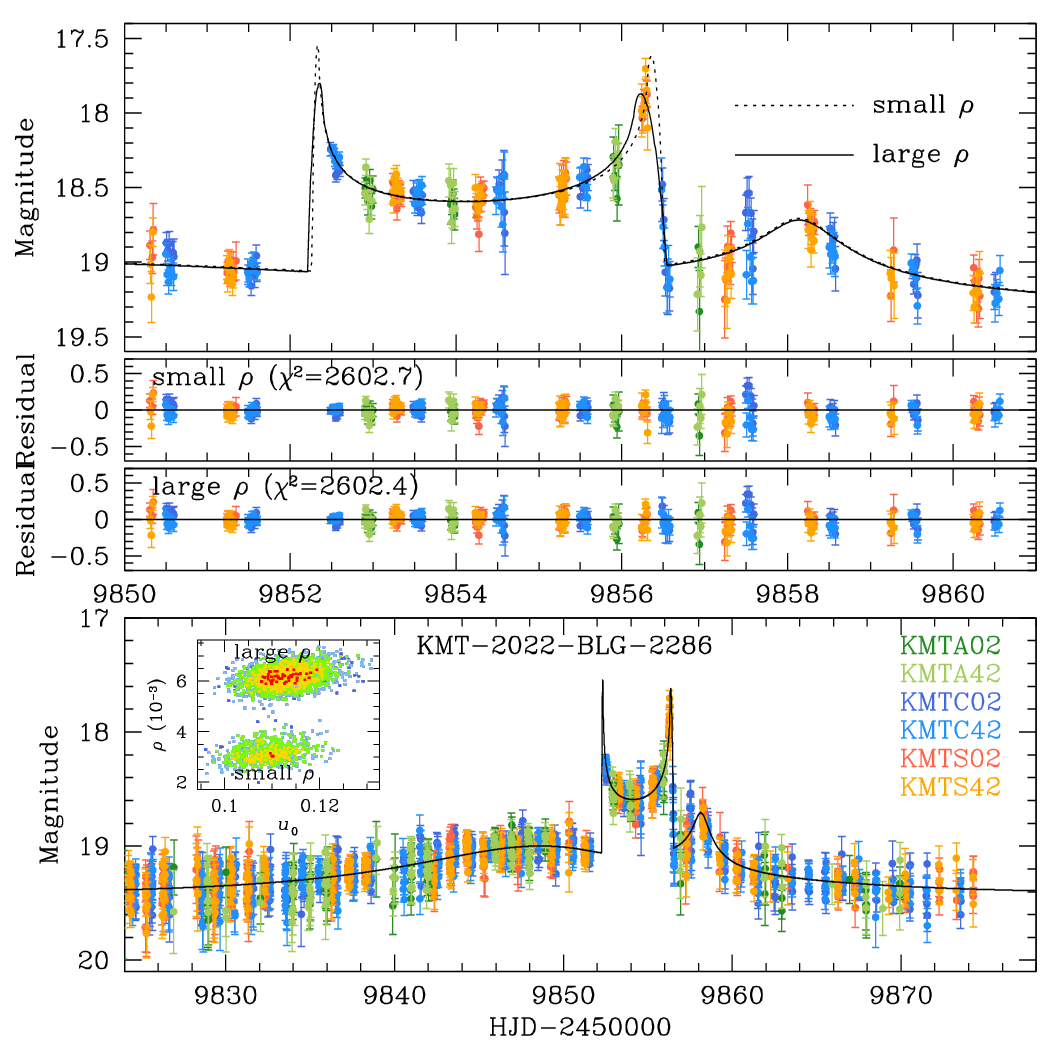}
\caption{
Lensing light curve of KMT-2022-BLG-2286.  The notations used are identical to those employed in
Fig.~\ref{fig:one}.  The inset in the bottom panel indicates the locations of the small-$\rho$ 
and large-$\rho$ solutions in the scatter plot of MCMC points on the $u_0$--$\rho$ plane.  The 
color scheme is set up to represent points as red ($\leq 1\sigma$), yellow ($\leq 2\sigma$), 
green ($\leq 3\sigma$), and cyan ($\leq 4\sigma$).
}
\label{fig:three}
\end{figure}

\subsection{KMT-2022-BLG-2286}\label{sec:four-two}

The KMTNet group discovered the lensing-induced source flux magnification of the lensing event
KMT-2022-BLG-2286 on October 4, 2022, which corresponds to ${\rm HJD}^\prime\sim 9856$.  
Figure~\ref{fig:three} displays the light curve of the lensing event.  Inspection of the light 
curve reveals two anomalous features: caustic-crossing feature with two spikes at ${\rm HJD}^\prime
\sim 9852.2$ and 9856.4 and a weak bump at ${\rm HJD}^\prime\sim 9858.1$.  The last observation 
of the season was done on October 21, ${\rm HJD}^\prime\sim 9874.2$.  The event was very densely 
observed due to the source being located in the overlapping region of the two prime KMTNet fields, 
BLG02 and BLG42. Each field was observed with a 0.5-hour cadence, resulting in a combined cadence 
of 0.25 hours.

We have identified two solutions with nearly identical binary parameters, $(s, q) \sim (1.4, 
0.09)$.  Table~\ref{table:three} presents the full lensing parameters for both solutions. 
The primary disparity between them lies in the normalized source radius: $(3.12\pm 0.32)\times 
10^{-3}$ for one solution and $(6.09\pm 0.49)\times 10^{-3}$ for the other, which we label as 
the "small-$\rho$" and "large-$\rho$" solutions, respectively.  In the inset of the bottom panel 
of Figure~\ref{fig:three}, we present locations of the two local solutions in the scatter plot 
of points on the MCMC chain on the $u_0$--$\rho$ parameter plane.  The degeneracy between the 
two solutions is very severe, with the large-$\rho$ solution preferred by only $\Delta\chi^2 = 
0.3$.  We conducted separate analyses for each solution due to discrepancies in the angular 
Einstein radii resulting from the differing normalized source radii. This distinction is crucial 
because $\thetae$ serves as a critical observable constraining the physical properties of the lens.  
The two local solutions yield different values for the angular Einstein radius and the relative 
lens-source proper motion. Then, this degeneracy can be resolved by measuring the relative 
lens-source proper motion through future Adaptive Optic (AO) observations.
The upper panels of Figure~\ref{fig:three} display the model curves and residuals for the two 
degenerate solutions. While subtle distinctions are evident in the model light curve during the 
caustic exit, pinpointing a definitive solution proves challenging due to data uncertainties in 
this region. 

\begin{figure}[t]
\includegraphics[width=\columnwidth]{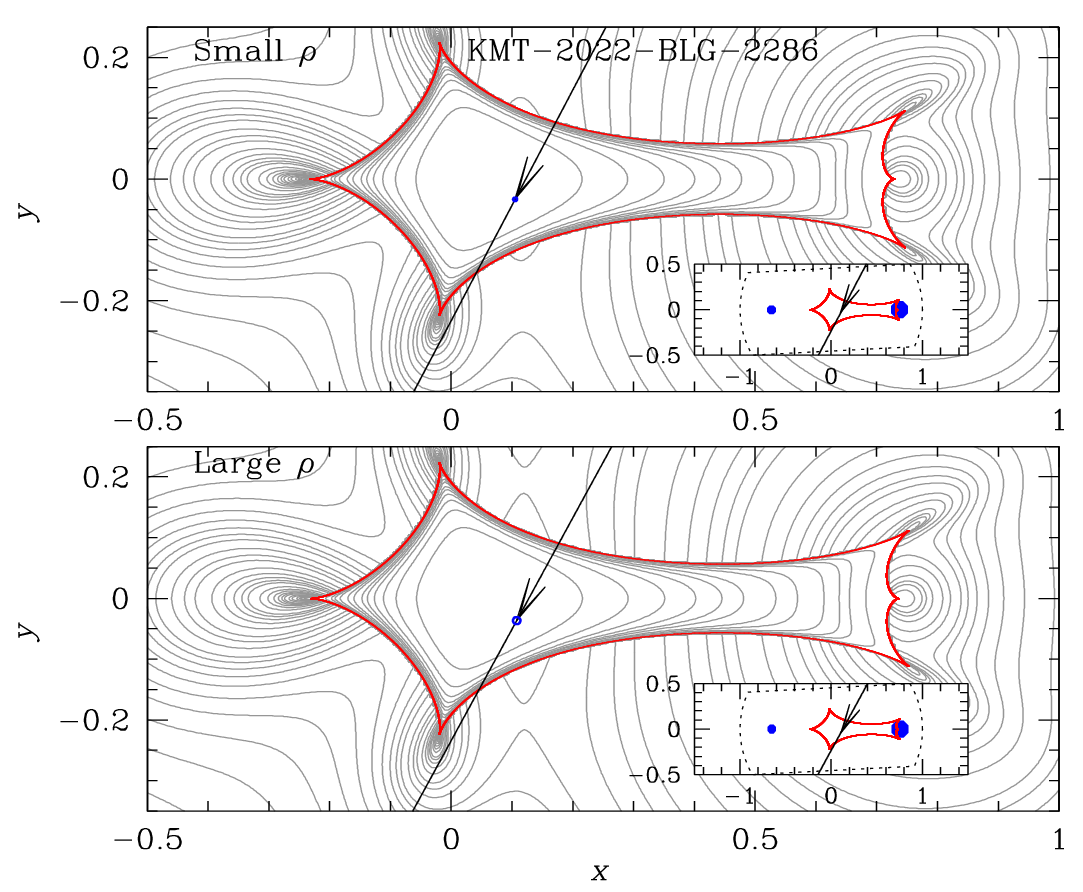}
\caption{
Lens-system configuration of KMT-2022-BLG-2286. The upper and lower panels show configurations of 
the "small-$\rho$" and "large-$\rho$" solutions, respectively. In each panel, the small empty circle 
on the source trajectory indicates the scaled size of the source.
}
\label{fig:four}
\end{figure}

\begin{table}[t]
\caption{Lensing parameters KMT-2022-BLG-2286.\label{table:three}}
\begin{tabular*}{\columnwidth}{@{\extracolsep{\fill}}lllcc}
\hline\hline
\multicolumn{1}{c}{Parameter}          &
\multicolumn{1}{c}{Small $\rho$}       &
\multicolumn{1}{c}{Large $\rho$}       \\
\hline
$\chi^2$                &  $2602.7            $     &   $2602.4             $   \\
$t_0$ (HJD$^\prime$)    &  $9854.695 \pm 0.056$     &   $9854.7088 \pm 0.056$   \\
$u_0$                   &  $0.1085 \pm 0.0052 $     &   $0.1120 \pm 0.0052  $   \\
$\te$ (days)            &  $14.69 \pm 0.34    $     &   $14.70 \pm 0.35     $   \\
$s$                     &  $1.4036 \pm 0.0060 $     &   $1.4096 \pm 0.0065  $   \\
$q$                     &  $0.0896 \pm 0.0063 $     &   $0.0915 \pm 0.0066  $   \\
$\alpha$ (rad)          &  $5.198 \pm 0.016   $     &   $5.205 \pm 0.016    $   \\
$\rho$ ($10^{-3}$)      &  $3.12 \pm 0.32     $     &   $6.09 \pm 0.49      $   \\
\hline
\end{tabular*}
\end{table}

Figure~\ref{fig:four} illustrates the configurations of the lens systems corresponding to the two 
solutions. As expected, given the similarity in their lensing parameters, the configurations closely 
resemble each other. We denote the source size, scaled to the caustic size, as an empty circle on 
the source trajectory. In both solutions, the lens forms a single resonant caustic elongated along 
the binary axis, with the companion positioned on the left side in the map. The source traverses 
diagonally through the region between the primary and companion lens components of the caustic, 
resulting in the observation of two caustic spikes. The bump occurring at ${\rm HJD}^\prime \sim 
9858.1$ corresponds to the moment when the source approaches the lower left cusp of the caustic.

\begin{table}[b]
\caption{Lensing parameters of KMT-2023-BLG-0201.\label{table:four}}
\begin{tabular*}{\columnwidth}{@{\extracolsep{\fill}}llll}
\hline\hline
\multicolumn{1}{c}{Parameter}   &
\multicolumn{1}{c}{Value}        \\
\hline
$\chi^2$                &  $1916.3             $  \\
$t_0$ (HJD$^\prime$)    &  $10048.980 \pm 0.011$  \\
$u_0$                   &  $0.3583 \pm 0.0012  $  \\
$\te$ (days)            &  $9.995 \pm 0.015    $  \\
$s$                     &  $1.10201 \pm 0.00058$  \\
$q$                     &  $0.2030 \pm 0.0020  $  \\
$\alpha$ (rad)          &  $3.4596 \pm 0.0012  $  \\
$\rho$ ($10^{-3}$)      &  $ 22.29 \pm 0.19    $  \\
\hline
\end{tabular*}
\end{table}

\subsection{KMT-2023-BLG-0201}\label{sec:four-three}

\begin{figure}[t]
\includegraphics[width=\columnwidth]{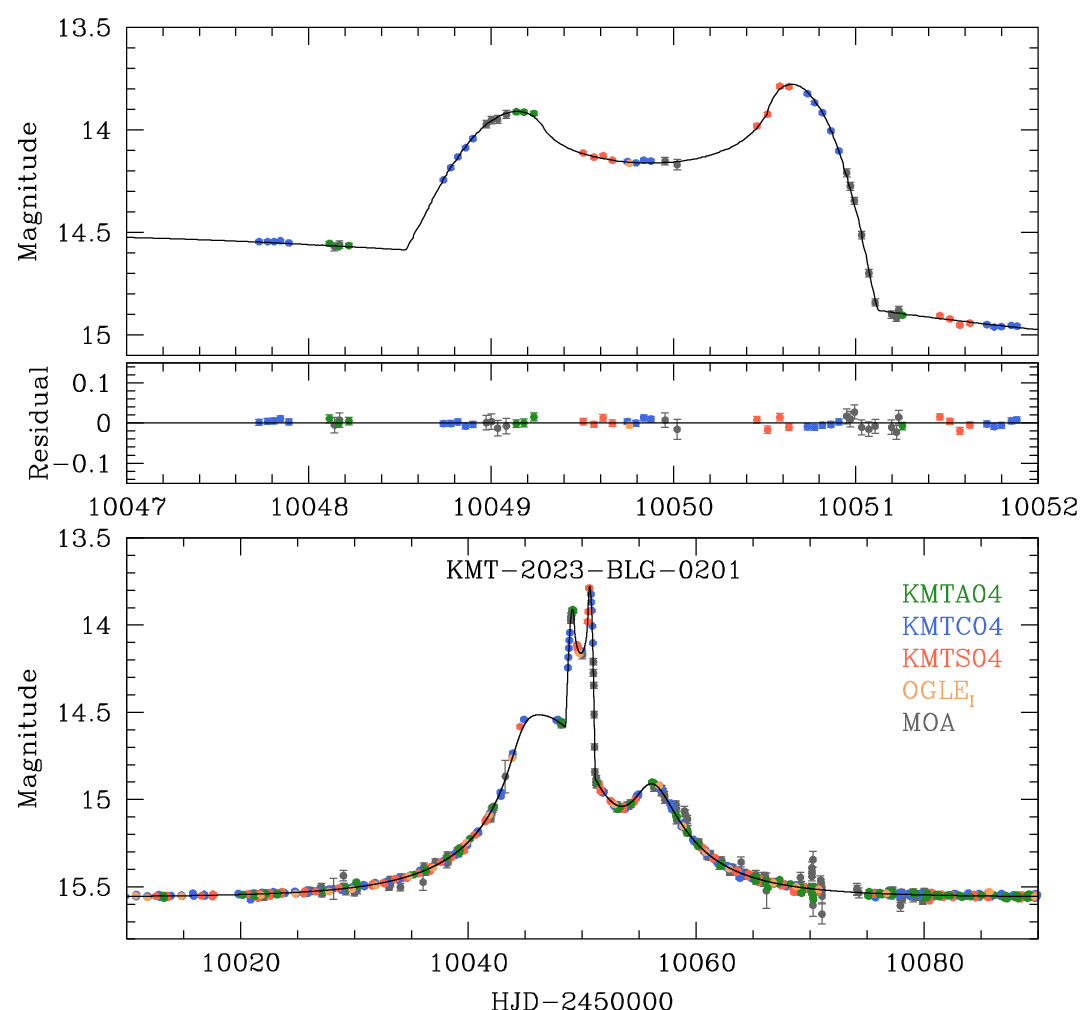}
\caption{
Light curve of the lensing event KMT-2023-BLG-0201.
}
\label{fig:five}
\end{figure}

The brightening of the source star in KMT-2023-BLG-0201 commenced early in the 2023 season, and 
the KMTNet survey identified the gravitational lensing nature of this light variation on March 29, 
2023 (${\rm HJD}^\prime \sim 10013.2$). The event was also observed by the OGLE and MOA groups, 
which labeled it OGLE-2023-BLG-0263 and MOA-2023-BLG-102, respectively. Following its detection, 
the event exhibited a sequence of anomalies, including a weak bump around ${\rm HJD}^\prime \sim 
10045$, two distinct caustic spikes at ${\rm HJD}^\prime \sim 10049.1$ and ${\rm HJD}^\prime \sim 
10053.2$, and another bump centered around ${\rm HJD}^\prime \sim 10056$. Figure~\ref{fig:five} 
presents the light curve of the event, combining data from the three lensing surveys.

Modeling the light curve resulted in a unique solution with binary parameters of $(s, q) \sim 
(1.1, 0.20)$ and an event timescale of $\te \sim 10$ days. Despite the mass ratio being higher 
than the basic threshold of $q_{\rm max} = 0.1$, the event was selected as a BD candidate due to 
its short timescale. The full lensing parameters are detailed in Table~\ref{table:four}. The 
caustic spikes exhibit pronounced finite source size effects, resulting in a rounded shape during 
caustic crossings. Both caustic-crossing segments of the light curve were densely resolved by the 
combined data sets, leading to a precisely measured normalized source radius of $\rho = (22.29 
\pm 0.19) \times 10^{-3}$.

\begin{figure}[t]
\includegraphics[width=\columnwidth]{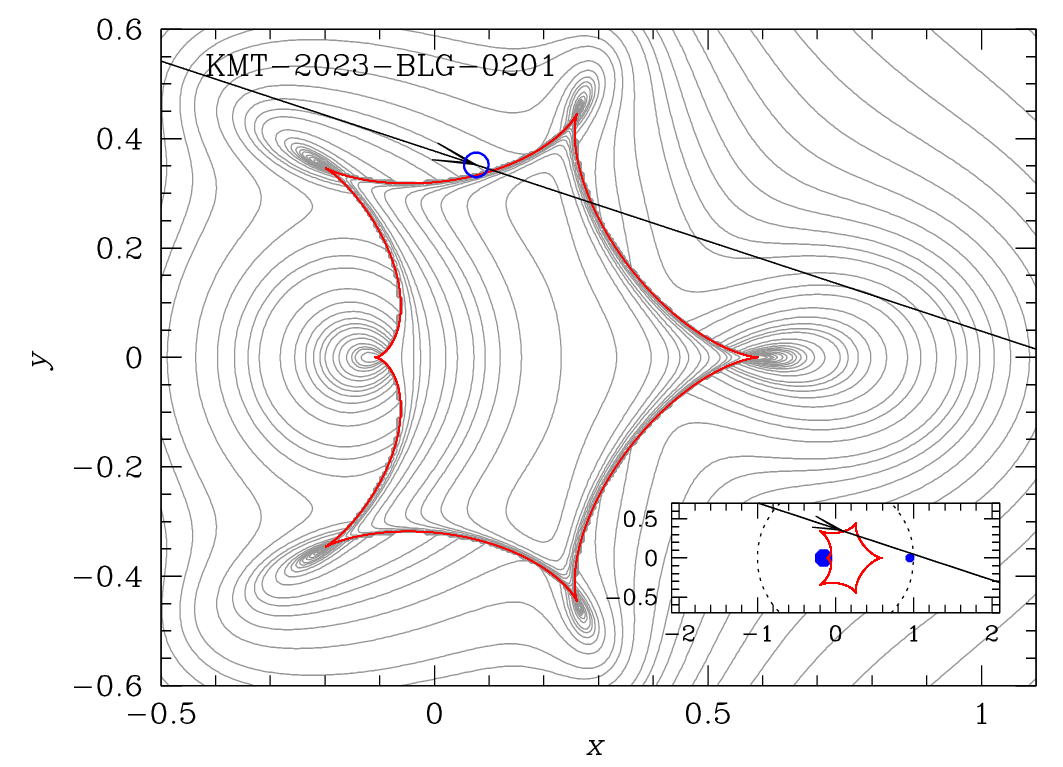}
\caption{
Configuration of the lens system KMT-2023-BLG-0201.
}
\label{fig:six}
\end{figure}

The lens-system configuration for the event is shown in Figure~\ref{fig:six}. As with the previous 
two events, the binary lens forms a resonant caustic. The source traversed the upper two folds of 
the caustic at an angle, resulting in the two observed caustic spikes. Before entering the caustic, 
the source approached the upper left cusp, generating the initial weak bump around ${\rm HJD}^\prime 
\sim 10045$. After exiting the caustic, it neared another cusp on the right side of the binary axis, 
causing a second bump around ${\rm HJD}^\prime \sim 10056$.

\subsection{KMT-2023-BLG-0601}\label{sec:four-four}

\begin{figure}[t]
\includegraphics[width=\columnwidth]{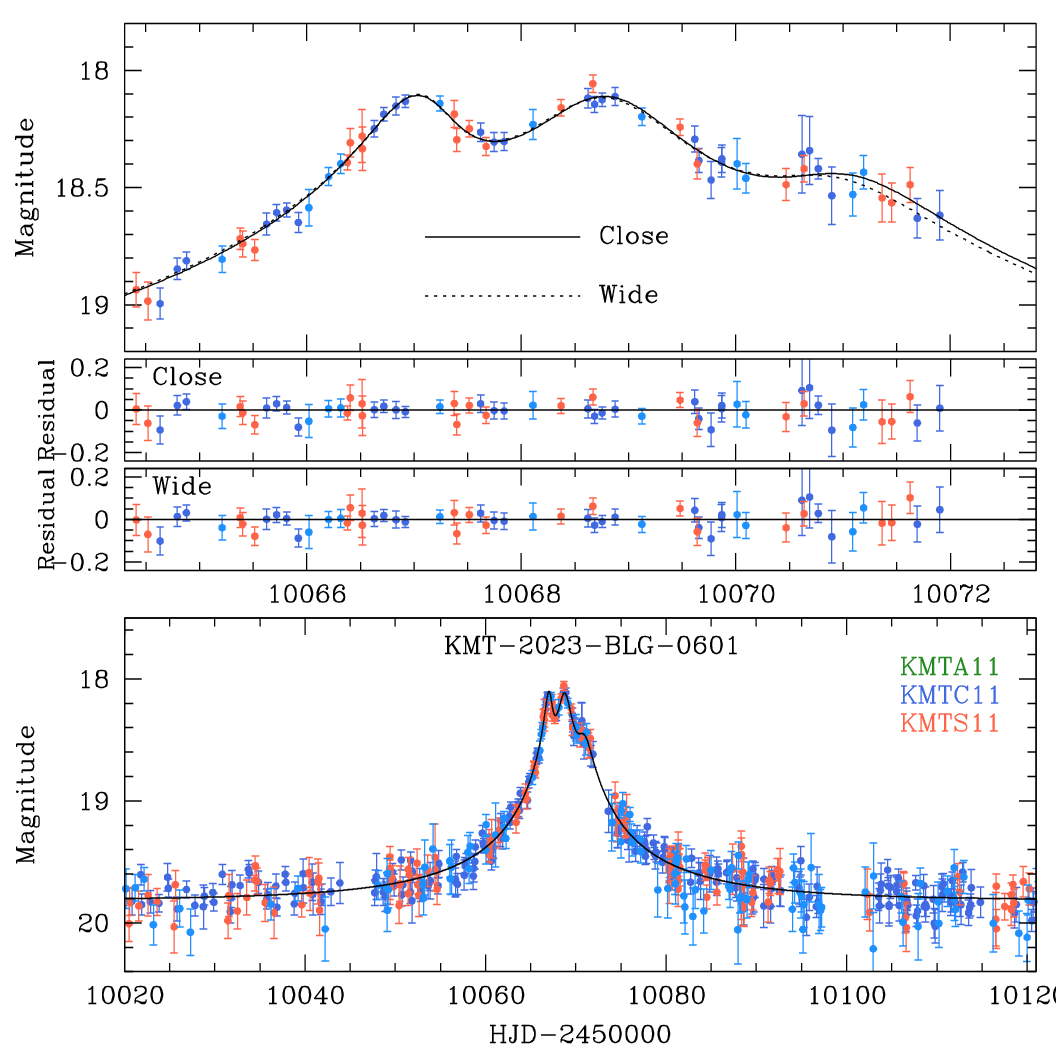}
\caption{
Light curve of the lensing event KMT-2023-BLG-0601.
}
\label{fig:seven}
\end{figure}

The lensing event KMT-2023-BLG-0601 was detected on April 28, 2023 (${\rm HJD}^\prime \sim 10063$), 
about five days before its peak. It was exclusively observed by the KMTNet group with a cadence of 
1.0 hour. Figure~\ref{fig:seven} presents the light curves constructed by combining data from the 
three KMTNet telescopes. The peak region of the light curve features three successive bumps, each 
separated by approximately two days. These bumps likely result from the source approaching close 
to the caustics near the primary lens. An anomaly consisting of three closely spaced bumps can 
occur when a caustic is skewed, with its cusps located on one side of the caustic produced by a 
binary lens. Such a skewed caustic can be formed by a binary lens with a small mass ratio between 
its components, as demonstrated in the cases of OGLE-2005-BLG-071Lb \citep{Udalski2005} and
KMT-2018-BLG-0885 \citep{Han2023a}.

\begin{table}[t]
\caption{Lensing parameters of KMT-2023-BLG-0601.\label{table:five}}
\begin{tabular*}{\columnwidth}{@{\extracolsep{\fill}}llll}
\hline\hline
\multicolumn{1}{c}{Parameter   }   &
\multicolumn{1}{c}{Small $\rho$}   &
\multicolumn{1}{c}{Large $\rho$}   \\
\hline
$\chi^2$                &   $930.9              $   &  $931.1              $ \\
$t_0$ (HJD$^\prime$)    &   $10068.687 \pm 0.036$   &  $10068.679 \pm 0.039$ \\
$u_0$                   &   $0.0615 \pm 0.0075  $   &  $0.0638 \pm 0.0059  $ \\
$\te$ (days)            &   $23.02 \pm 2.93     $   &  $23.24 \pm 1.70     $ \\
$s$                     &   $0.557 \pm 0.013    $   &  $1.719 \pm 0.046    $ \\
$q$                     &   $0.0643 \pm 0.0078  $   &  $0.0674 \pm 0.0063  $ \\
$\alpha$ (rad)          &   $1.426 \pm 0.016    $   &  $1.42 \pm 0.018     $ \\
$\rho$ ($10^{-3}$)      &   --                      &  --                    \\
\hline
\end{tabular*}
\end{table}

Detailed modeling of the light curve resulted in two sets of solutions. The binary parameters for 
these solutions are $(s, q)_{\rm close} \sim (0.56, 0.064)$ and $(s, q)_{\rm wide} \sim (1.72, 
0.067)$.  The full lensing parameters for both solutions, along with their $\chi^2$ values, are 
listed in Table~\ref{table:five}. Figure~\ref{fig:seven} presents the model curves and residuals 
for the two solutions. The difference in $\chi^2$ between them is very minor, $\Delta\chi^2 = 0.2$, 
indicating that the solutions are highly degenerate. The similarity between the model curves arises 
from the well-known close-wide degeneracy \citep{Griest1998, Dominik1999, An2005}, as shown by the 
binary separations being related by $s_{\rm close} \sim 1/s_{\rm wide}$ and their similar mass 
ratios.

\begin{figure}[t]
\includegraphics[width=\columnwidth]{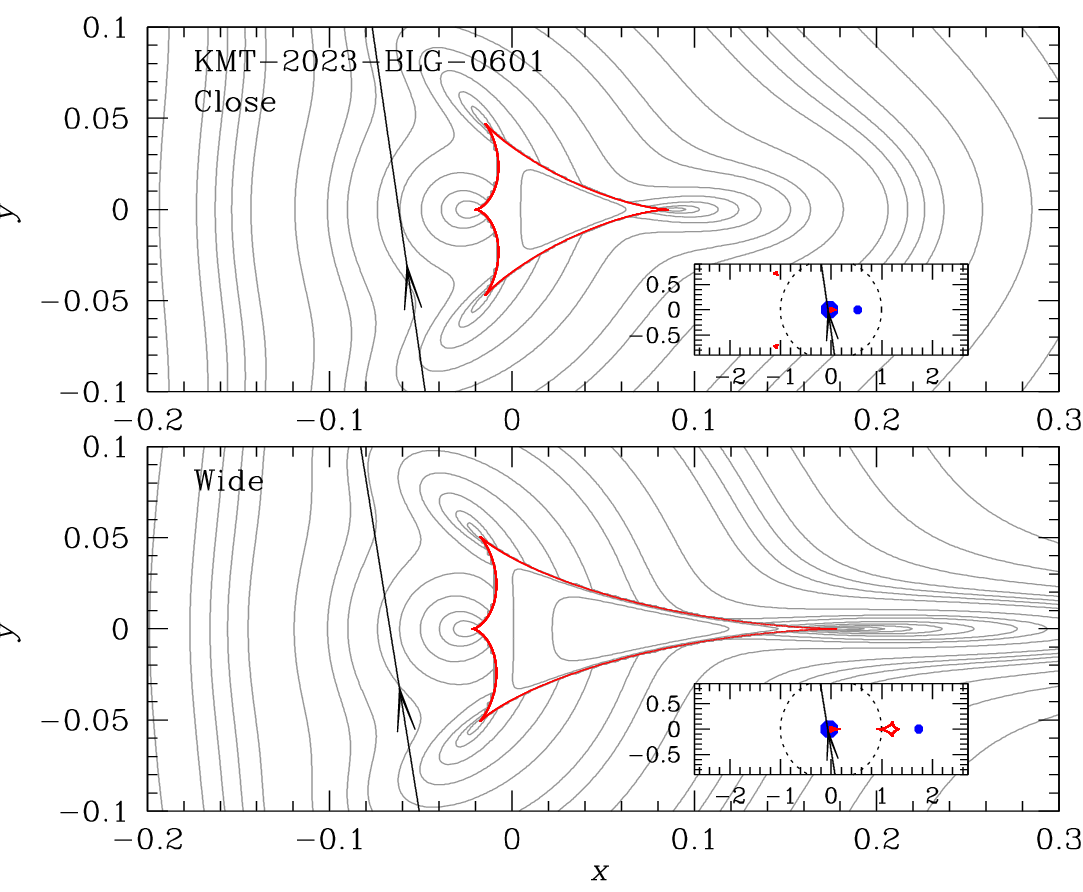}
\caption{
Configuration of the lens system KMT-2023-BLG-0601.
}
\label{fig:eight}
\end{figure}

Figure~\ref{fig:eight} displays the configurations of the lens system for the close (upper panel) 
and wide (lower panel) solutions. In both solutions, the source traverses the region behind the 
central caustic induced by a low-mass companion. As anticipated from the anomaly pattern, three 
of the four cusps of the central caustic are positioned on one side, resulting in the generation 
of weak bumps as the source approaches these cusps. In planetary lensing scenarios, a similar 
source trajectory would lead to an anomaly pattern where the middle bump is suppressed due to 
the relatively weaker middle cusp compared to the others \citep{Han2008}. Consequently, planetary 
scenarios of the anomaly were excluded. The normalized source radius could not be determined because 
the separation between the source and caustic during the anomalies was too large to induce noticeable 
finite source effects.

\subsection{KMT-2023-BLG-1684}\label{sec:four-five}

\begin{figure}[t]
\includegraphics[width=\columnwidth]{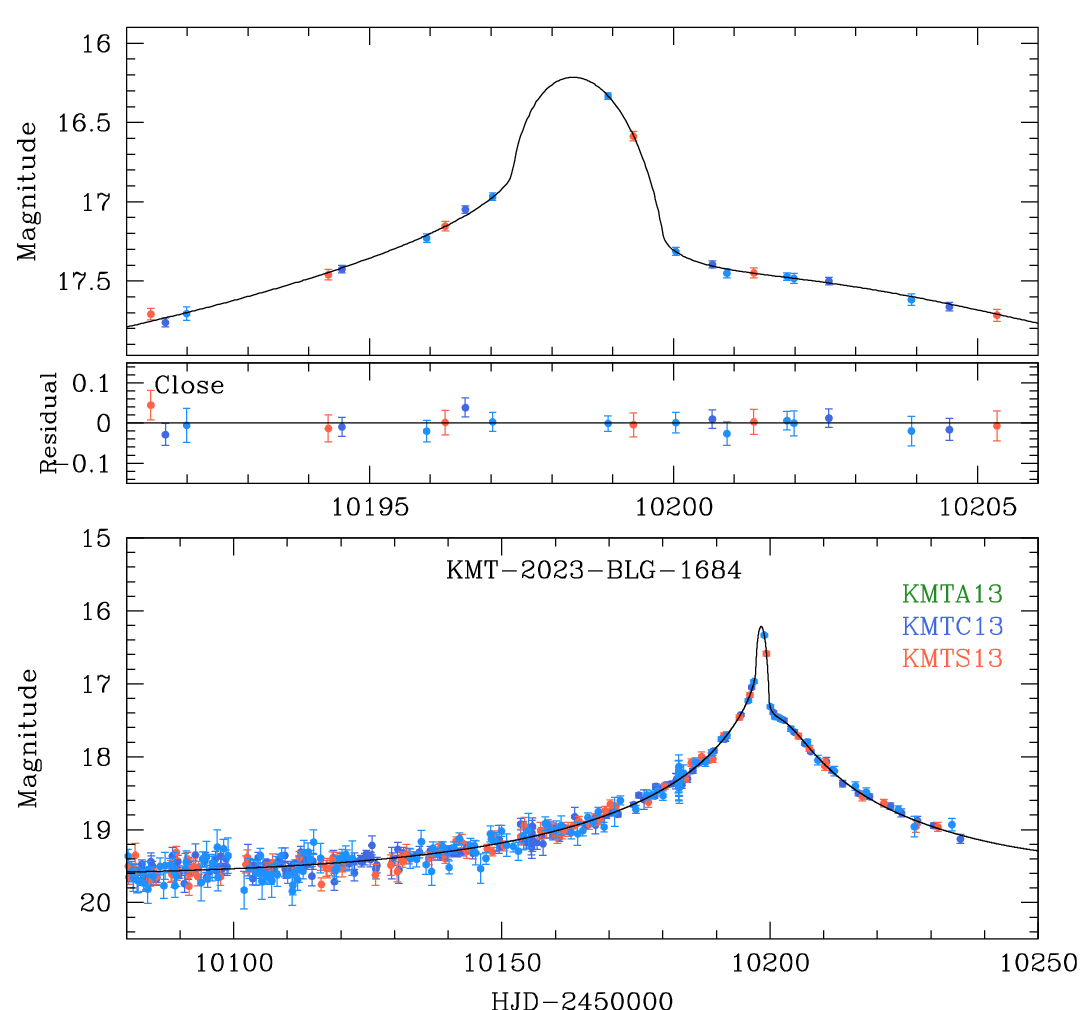}
\caption{
Light curve of the lensing event KMT-2023-BLG-1684.	
}
\label{fig:nine}
\end{figure}

The lensing-induced magnification of the source flux in the event KMT-2023-BLG-1684 commenced in 
the middle of the 2023 bulge season and persisted beyond its conclusion. Figure~\ref{fig:nine} 
illustrates the light curve of the event.  The light curve was well characterized despite a 
relatively lower cadence (5 hours) and a very high extinction ($A_I = 5$~mag) toward the field.  
The KMTNet group detected the event early in its magnification phase on July 17, 2023 (${\rm HJD}^\prime 
\sim 10143$).  The light curve peaked around ${\rm HJD}^\prime =10198$, during which it exhibited an 
anomaly characterized by a positive deviation lasting approximately three days. Another notable 
aspect of the anomaly is the significant disparity in lensing magnifications observed before and 
after the major anomaly.

\begin{table}[b]
\caption{Best-fit lensing parameters of KMT-2023-BLG-1684.\label{table:six}}
\begin{tabular*}{\columnwidth}{@{\extracolsep{\fill}}llll}
\hline\hline
\multicolumn{1}{c}{Parameter}   &
\multicolumn{1}{c}{Value}        \\
\hline
$\chi^2$                &  $679.9              $  \\
$t_0$ (HJD$^\prime$)    &  $10199.300 \pm 0.212$  \\
$u_0$                   &  $0.0995 \pm 0.0067  $  \\
$\te$ (days)            &  $64.27 \pm 2.97     $  \\
$s$                     &  $0.693 \pm 0.019    $  \\
$q$                     &  $0.068 \pm 0.010    $  \\
$\alpha$ (rad)          &  $0.534 \pm 0.056    $  \\
$\rho$ ($10^{-3}$)      &  $18.4 \pm 1.53      $  \\
\hline
\end{tabular*}
\end{table}

\begin{figure}[t]
\includegraphics[width=\columnwidth]{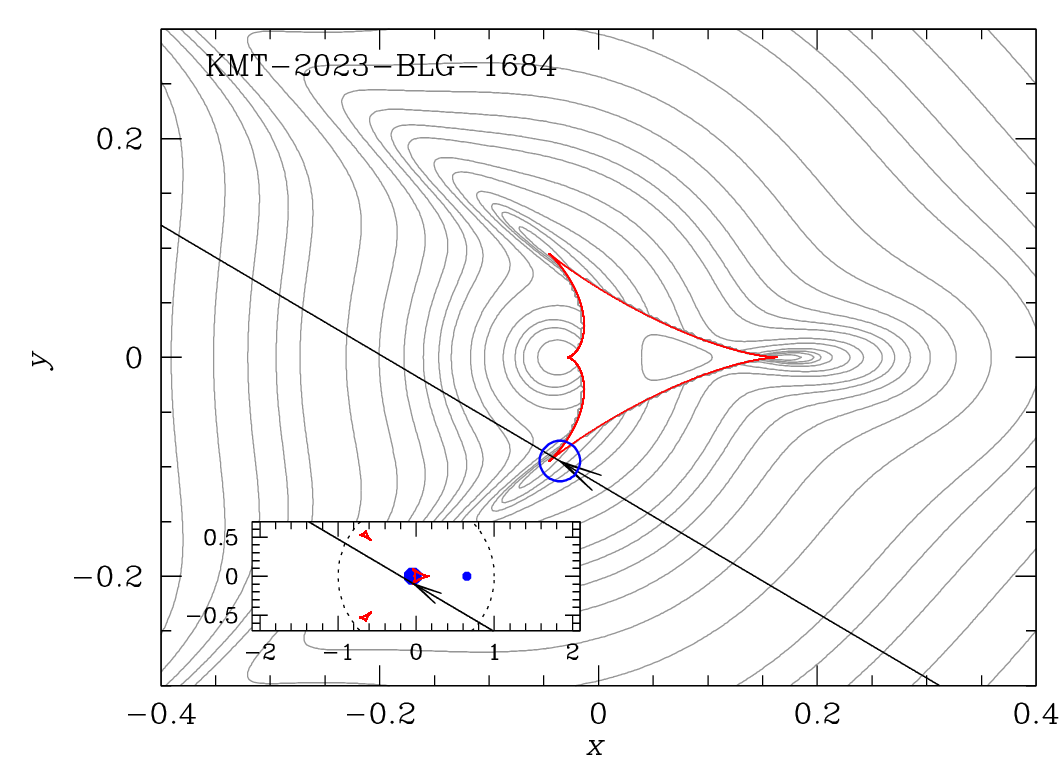}
\caption{
Lens-system configuration of the lensing event KMT-2023-BLG-1684.
}
\label{fig:ten}
\end{figure}

The analysis of the anomaly revealed a unique solution with binary parameters $(s, q) \sim (0.69, 
0.068)$, indicating that the lens companion falls within the brown dwarf mass range. The full 
lensing parameters for this solution are listed in Table~\ref{table:six}. The event's measured 
timescale, $\te \sim 64$ days, is relatively long, prompting additional modeling to consider 
higher-order effects.  However, accurately determining higher-order lensing parameters proved 
challenging, mainly because the declining part of the light curve had incomplete coverage.  
Nevertheless, the normalized source radius, $\rho = (18.4 \pm 1.53) \times 10^{-3}$, was precisely 
determined from the analysis of the anomaly, which was significantly influenced by finite source 
effects.

Figure~\ref{fig:ten} shows the arrangement of the source trajectory with respect to the caustic. 
The companion lying inside the Einstein ring induces two sets of caustics: one near the position 
of the primary and two peripheral caustics situated off the binary axis on the opposite side of 
the companion. The source crossed the lower tip of the central caustic, generating the anomaly 
near the peak of the light curve. The difference in magnification before and after crossing the 
caustic is attributed to negative magnification excess in the region behind the central caustic.

\subsection{KMT-2023-BLG-1743}\label{sec:four-six}

\begin{figure}[t]
\includegraphics[width=\columnwidth]{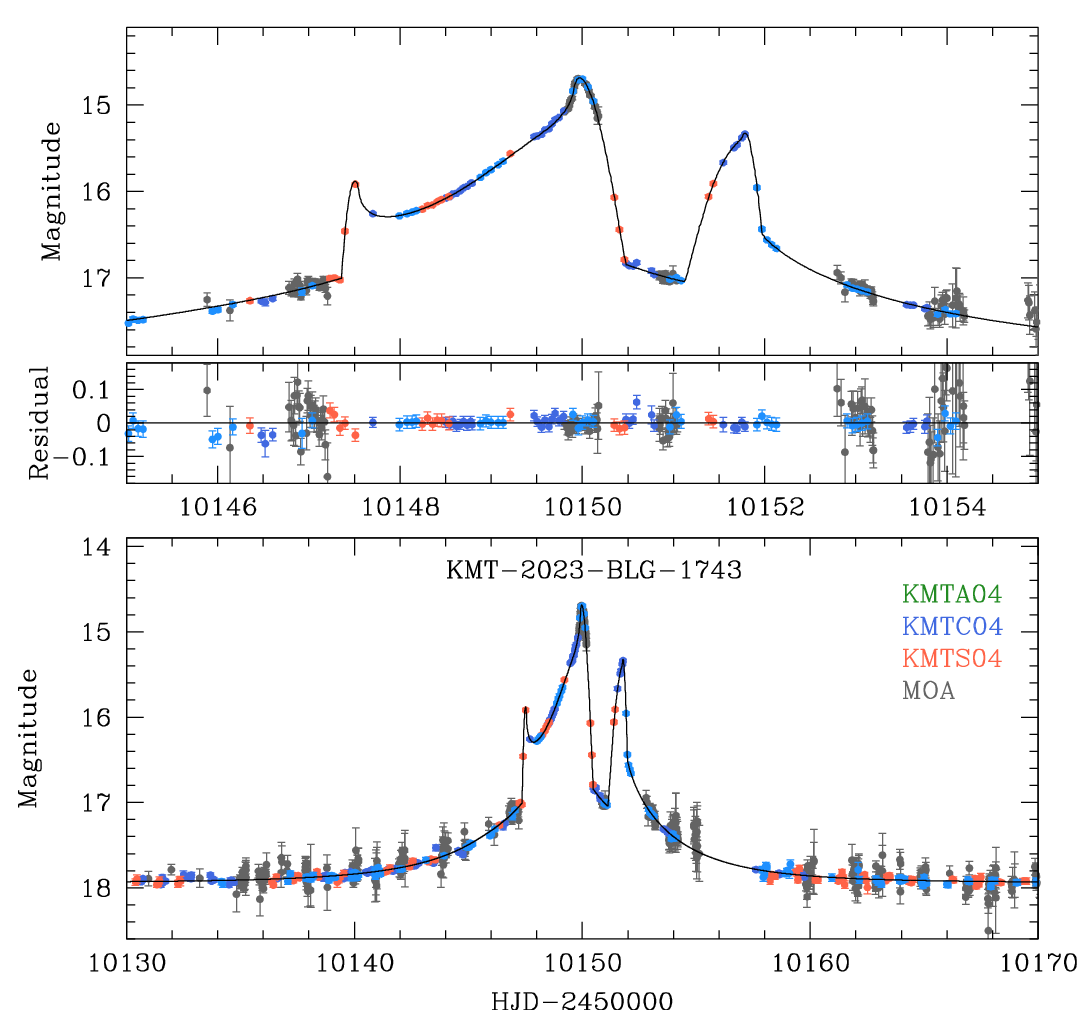}
\caption{
Light curve of the lensing event KMT-2023-BLG-1743.
}
\label{fig:eleven}
\end{figure}

The KMTNet group initially detected the lensing event KMT-2023-BLG-1743 on July 19, 2023 
(${\rm HJD}^\prime \sim 10145$), and two days later, the MOA group confirmed it, designating 
it as MOA-2023-BLG-324. Figure~\ref{fig:eleven} illustrates the light curve of the event, 
which exhibited two consecutive anomaly sets of caustic-crossing features. The first set 
displayed two distinct caustic peaks at ${\rm HJD}^\prime \sim 10147.3$ and ${\rm HJD}^\prime 
\sim 10149.9$, while the second set exhibited a single peak centered at ${\rm HJD}^\prime \sim 
10151.8$. The asymmetric pattern between the caustic peaks of the first set suggests that the 
source approached a fold of the caustic asymptotically. The absence of a U-shaped pattern 
between the rising and falling parts of the latter anomaly set implies that the source crossed 
the tip of the caustic.

\begin{table}[t]
\caption{Best-fit lensing parameters of KMT-2023-BLG-1743.\label{table:seven}}
\begin{tabular*}{\columnwidth}{@{\extracolsep{\fill}}llll}
\hline\hline
\multicolumn{1}{c}{Parameter}   &
\multicolumn{1}{c}{Value}       \\
\hline
$\chi^2$                &  $2118.3               $  \\
$t_0$ (HJD$^\prime$)    &  $60149.3030 \pm 0.0099$  \\
$u_0$                   &  $-0.0023 \pm 0.0016   $  \\
$\te$ (days)            &  $5.876 \pm 0.049      $  \\
$s$                     &  $1.0136 \pm 0.0008    $  \\
$q$                     &  $0.1770 \pm 0.0031    $  \\
$\alpha$ (rad)          &  $5.2266 \pm 0.0070    $  \\
$\rho$ ($10^{-3}$)      &  $14.49 \pm 0.25       $  \\
\hline
\end{tabular*}
\end{table}

\begin{table*}[t]
\caption{Source parameters of KMT-2033-BLG-0412, KMT-2022-BLG-2286, and KMT-2023-BLG-0201.\label{table:eight}}
\begin{tabular}{lllll}
\hline\hline
\multicolumn{1}{c}{Quantity}            &
\multicolumn{1}{c}{KMT-2022-BLG-0412}   &
\multicolumn{1}{c}{KMT-2022-BLG-2286}   &
\multicolumn{1}{c}{KMT-2023-BLG-0201}   \\
\hline
$(V-I, I)$                  &  $(1.830\pm 0.096, 20.581\pm 0.018)$  &  $(2.784\pm 0.076, 20.055\pm 0.015)$    &  $(2.049\pm 0.049, 15.717\pm 0.002)$   \\
$(V-I, I)_{\rm RGC}$        &  $(1.761, 15.390)                  $  &  $(2.828, 16.455)                  $    &  $(2.205, 16.175)                  $   \\
$(V-I, I)_{{\rm RGC},0}$    &  $(1.060, 14.292)                  $  &  $(1.060, 14.389)                  $    &  $(1.060, 14.433)                  $   \\
$(V-I, I)_{0}$              &  $(1.129\pm 0.104, 19.483\pm 0.027)$  &  $(1.016\pm 0.086, 17.989\pm 0.025)$    &  $(0.904\pm 0.064, 13.975\pm 0.020)$   \\
 Source type                &   K3.5V                               &   K3V                                   &   K2III                                \\
\hline
\end{tabular}
\end{table*}

\begin{table*}[t]
\caption{Source parameters of KMT-2023-BLG-0601, KMT-2023-BLG-1684, and KMT-2023-BLG-1743.\label{table:nine}}
\begin{tabular}{lllll}
\hline\hline
\multicolumn{1}{c}{Quantity}            &
\multicolumn{1}{c}{KMT-2023-BLG-0601}   &
\multicolumn{1}{c}{KMT-2023-BLG-1684}   &
\multicolumn{1}{c}{KMT-2023-BLG-1743}   \\
\hline
$(V-I, I)$                  &  $(1.789\pm 0.061, 22.320\pm 0.012)$  &  $(3.718\pm 0.139, 19.801\pm 0.056)$    &   $(2.198\pm 0.003, 18.063\pm 0.001)$   \\
$(V-I, I)_{\rm RGC}$        &  $(2.151, 17.717)                  $  &  $(3.644, 20.045)                  $    &   $(2.287, 16.144)                  $   \\
$(V-I, I)_{{\rm RGC},0}$    &  $(1.060, 14.576)                  $  &  $(1.060, 14.620)                  $    &   $(1.060, 14.576)                  $   \\
$(V-I, I)_{0}$              &  $(0.698\pm 0.073, 19.179\pm 0.023 $  &  $(1.56\pm 0.06, 14.375\pm 0.056)$      &   $(0.971\pm 0.040, 16.495\pm 0.020)$   \\
 Source type                &   G2V                                 &   K7III                                 &    K3III                                \\
\hline
\end{tabular}
\end{table*}

\begin{table*}[t]
\caption{Angular source radius, Einstein radius, and relative proper motion.\label{table:ten}}
\begin{tabular}{lllll}
\hline\hline
\multicolumn{1}{c}{Event}                    &
\multicolumn{1}{c}{$\theta_*$ ($\mu$as)}     &
\multicolumn{1}{c}{$\thetae$ (mas)}          &
\multicolumn{1}{c}{$\mu$ (mas/yr)}           \\
\hline
 KMT-2022-BLG-0412                &  $0.637 \pm 0.080$   &   $ > 0.16        $   &  $> 1.15$             \\
 KMT-2022-BLG-2286 (small $\rho$) &  $1.12 \pm 0.13  $   &   $0.361 \pm 0.055$   &  $8.97 \pm 0.14  $    \\
 \hskip95pt (large $\rho$)        &  $1.12 \pm 0.13  $   &   $0.185 \pm 0.023$   &  $4.59 \pm 0.56  $    \\
 KMT-2023-BLG-0201                &  $6.33 \pm 0.60  $   &   $0.282 \pm 0.027$   &  $10.30 \pm 0.97 $    \\
 KMT-2023-BLG-0601                &  $0.451 \pm 0.046$   &    --                 &   --                  \\
 KMT-2023-BLG-1684                &  $9.13 \pm 1.42  $   &   $0.50 \pm 0.09$     &  $2.82 \pm 0.51$      \\
 KMT-2023-BLG-1743                &  $2.18 \pm 0.18  $   &   $0.157 \pm 0.013$   &  $9.60 \pm 0.79  $    \\
\hline
\end{tabular}
\end{table*}

The analysis of the light curve revealed a unique solution with binary parameters 
$(s, q) \sim (1.01, 0.18)$. Despite the mass ratio exceeding 0.1, the event was classified as 
a BD candidate due to its short timescale ($\te \sim 5.9$ days). All four caustic peaks were 
resolved using the combined data, leading to a precisely determined value of the normalized 
source radius $\rho = (14.49 \pm 0.25) \times 10^{-3}$. The complete lensing parameters are 
listed in Table~\ref{table:seven}.

\begin{figure}[t]
\includegraphics[width=\columnwidth]{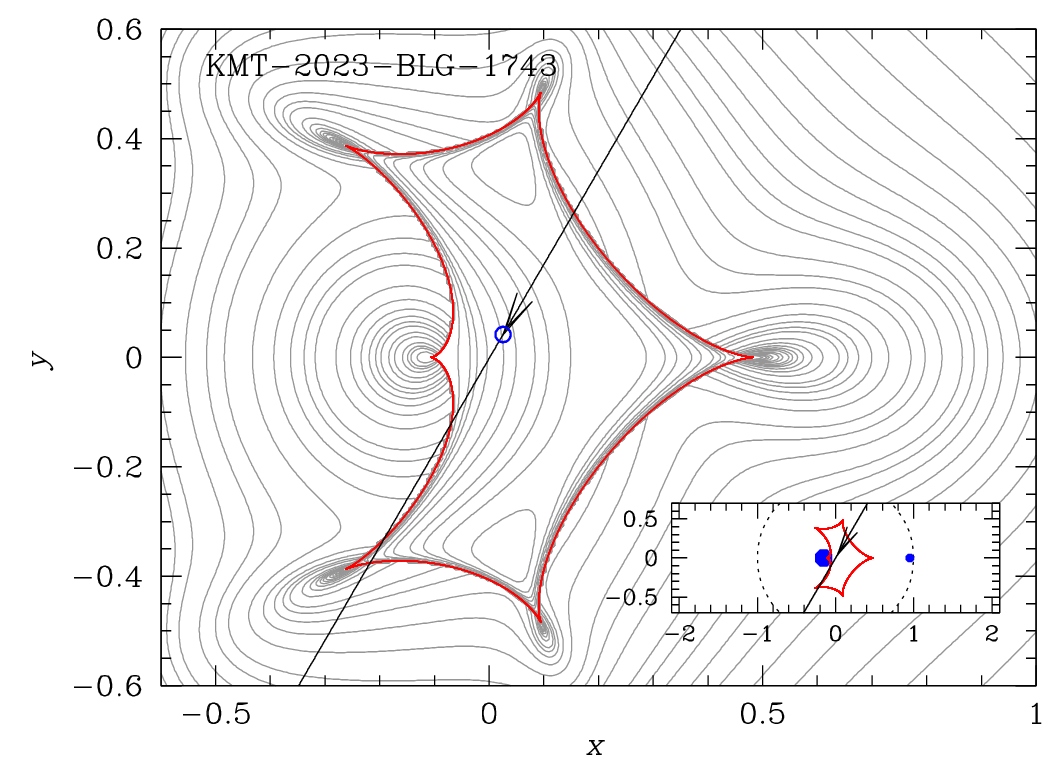}
\caption{
Lens-system configuration of the event KMT-2023-BLG-1743.
}
\label{fig:twelve}
\end{figure}

Figure~\ref{fig:twelve} illustrates the configuration of the lens system. The binary lens generated 
a resonant caustic composed of six folds. The source passed diagonally through the caustic, crossing 
these folds four times. Initially, the source entered the caustic by crossing the upper right fold 
and exited by traversing the lower left fold. This produced a classical caustic feature characterized 
by two spikes with a U-shaped trough region between them. The source then entered and exited the 
caustic again by moving through the area around the lower left cusp. The gap between this second 
entrance and exit was comparable to the size of the caustic, resulting in a single, merged anomaly 
feature.

\section{Source stars and angular Einstein radii}\label{sec:five}

In this section, we specify the source stars participating in the lensing events.  Besides offering 
a detailed overview of each event, defining the source star is essential for calculating the angular 
Einstein radius using the relationship
\begin{equation}
\thetae = {\thetae \over \rho}.
\label{eq1}
\end{equation}
The angular source radius, $\theta_*$, is inferred from the source type, while the normalized source 
radius is determined through modeling.

The source is specified by assessing its color and brightness following the methodology outlined by 
\citet{Yoo2004}. This method involves initially measuring the instrumental color and brightness of 
the source, which are then calibrated using the centroid of the red giant clump (RGC) in the 
color-magnitude diagram (CMD). The RGC centroid serves as a reference for this calibration due 
to its role as a standard candle, with its de-reddened color and magnitude for bulge stars well 
established in studies by \citet{Bensby2013} and \citet{Nataf2013}.

\begin{figure}[t]
\includegraphics[width=\columnwidth]{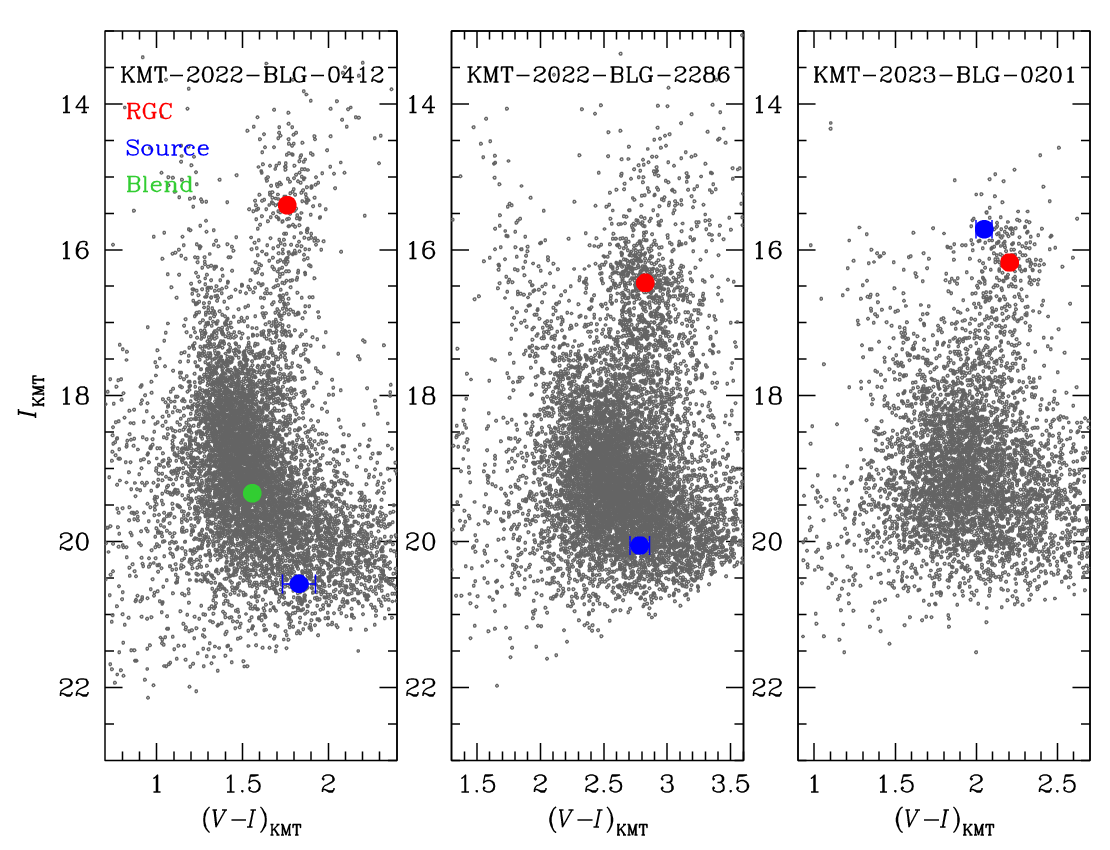}
\caption{
Color-magnitude diagrams of source stars for the lensing event KMT-2022-BLG-0412, KMT-2022-BLG-2286, 
and KMT-2023-BLG-0201. In each panel, small filled dots marked in red and blue represent the positions 
of red giant clump (RGC) centroid and source, respectively. For the event KMT-2022-BLG-0412, we 
additionally mark the position of the blend.
}
\label{fig:thirteen}
\end{figure}

\begin{figure}[t]
\includegraphics[width=\columnwidth]{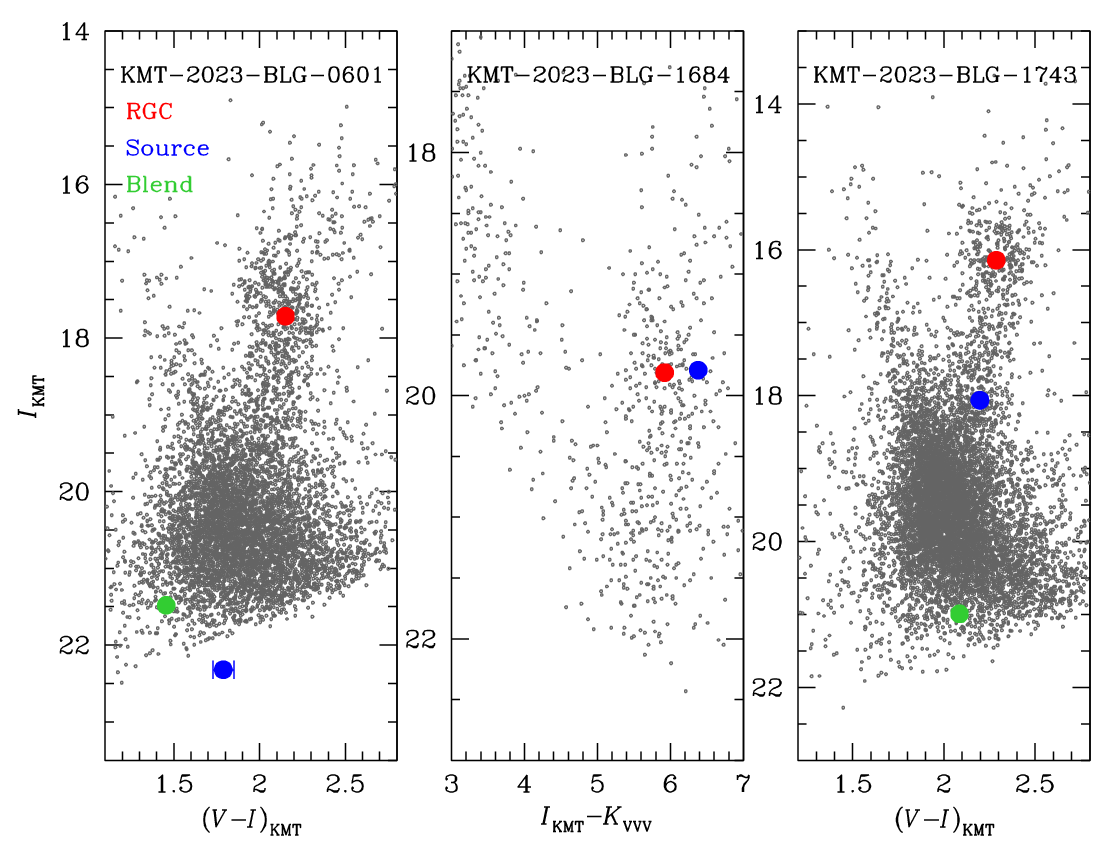}
\caption{
Color-magnitude diagrams of source stars for the lensing event KMT-2023-BLG-0601, KMT-2023-BLG-1684, 
and KMT-2023-BLG-1743.  Notations are same as those in Fig.~\ref{fig:thirteen}.  In the case of 
KMT-2023-BLG-1684, the color-magnitude diagram was compiled by aligning data from KMTNet 
observations with the VVV survey data.  
}
\label{fig:fourteen}
\end{figure}

In Figures~\ref{fig:thirteen} and \ref{fig:fourteen}, we present the source locations relative 
to the RGC centroids in the instrumental CMDs, constructed using KMTNet images processed with 
the pyDIA code \citep{Albrow2017}.  The instrumental source color and brightness, $(V-I, I)$, 
were determined by regressing the photometry data processed with the pyDIA code against the model. 
For events with measured blend colors, we also indicate the positions of the blends.  For 
KMT-2023-BLG-1684, measuring the $V-I$ source color was challenging due to the severe extinction 
toward the field, with $A_I = 5.41$.  In this case, we constructed $(I - K, I)$ CMD by matching 
KMT $I$-band pyDIA photometry to $K$-band VVV survey data \citep{Minniti2017}.  The source magnitude 
from the fit, $I=19.801\pm 0.056$, is identical to the baseline magnitude, leading us to conclude 
that the source is unblended in the $I$ band. Upon examining the nearest VVV entry, we found that 
the source is redder than the RGC centroid by $\Delta(I-K)=0.46$.

With the measured instrumental color and magnitude, the de-reddened values were estimated as
\begin{equation}
(V-I, I)_0 = (V-I, I)_{{\rm RGC},0} + \Delta(V-I, I), 
\label{eq2}
\end{equation}
where $(V-I, I)_{{\rm RGC},0}$ represent the de-reddened values of the RGC centroid and 
$\Delta(V-I, I)$ denote the offset of the source from the RGC centroid. In Tables~\ref{table:eight} 
and \ref{table:nine}, we list the values of $(V-I, I)$, $(V-I, I)_{\rm RGC}$, $(V-I, I)_{{\rm RGC},0}$, 
and $(V-I, I)_0$ together with the spectral types of the source stars. The source stars for 
KMT-2022-BLG-0412, KMT-2022-BLG-2286, and KMT-2023-BLG-0601 are main-sequence stars with spectral 
types K3.5, K3, and G2, respectively. On the other hand, the source stars for KMT-2023-BLG-0201, 
KMT-2023-BLG-1684, and KMT-2023-BLG-1743 are giant stars with spectral types K2, K7, and K3, 
respectively.  For KMT-2023-BLG-1684, we first converted the measured color offset $\Delta(I-K)$ 
into $\Delta(V-I)$ using the color-color relation of \citet{Bessell1988}, and then estimated the 
de-reddened color $\Delta(V-I)_0$.  We further verified the nature of the source by identifying 
its location in the $(J-K, I)$ CMD, which yielded a consistent result.

\begin{figure}[t]
\includegraphics[width=\columnwidth]{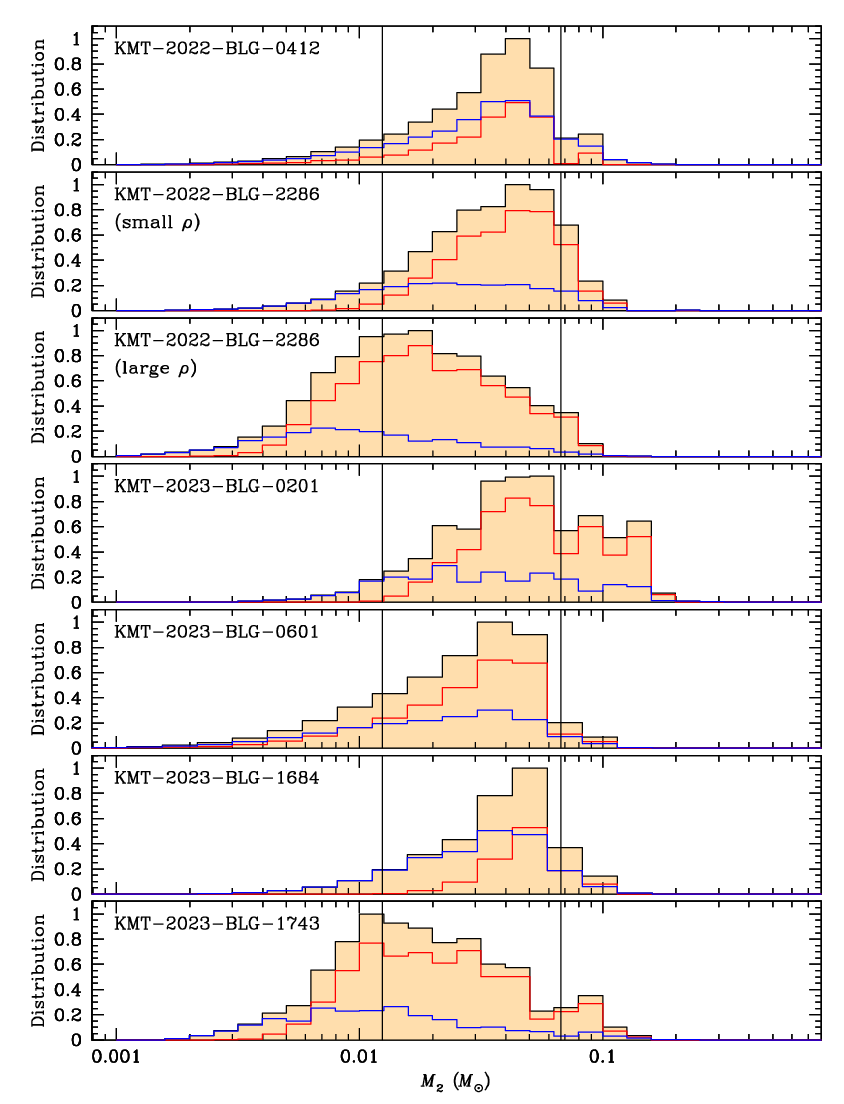}
\caption{
Bayesian posteriors for the mass of the lens companion. For KMT-2022-BLG-2286, two posteriors 
corresponding to the "small $\rho$" and "large $\rho$" solutions are presented. The two vertical 
lines in each panel represent the BD mass range, which is set as 13 to 80 times of the Jupiter. 
The red and blue distributions represent the contributions from the disk and bulge lens populations, 
respectively, while the black curve represents the combined contributions of both populations.
}
\label{fig:fifteen}
\end{figure}

To estimate the angular source radius, we first converted the measured $V-I$ color to $V-K$ using 
the color-color relation from \citet{Bessell1988}. We then estimated $\theta_*$ using the relation 
between the angular source radius and $V-K$ color provided by \citet{Kervella2004}. With the estimated 
value of $\theta_*$, the angular Einstein radius was determined using the relation in Eq.~(\ref{eq1}). 
The relative lens-source proper motion was estimated from the measured angular Einstein radius and 
event timescale as
\begin{equation}
\mu = {\thetae \over \te}.
\label{eq3}
\end{equation}
In Table~\ref{table:ten}, we list the estimated values of $\theta_*$, $\thetae$, and $\mu$ for the 
individual events. For KMT-2022-BLG-2286, which has two degenerate solutions with small and large 
$\rho$ values, we present two sets of $(\thetae, \mu)$ values corresponding to each solution.  For 
KMT-2022-BLG-0412, for which only the upper limit of $\rho$ is constrained, we provide the lower 
limits of $\thetae$ and $\mu$.  In the case of KMT-2023-BLG-0601, the normalized source radius 
could not be measured, and thus $(\thetae, \mu)$ values are not presented.

\begin{figure}[t]
\includegraphics[width=\columnwidth]{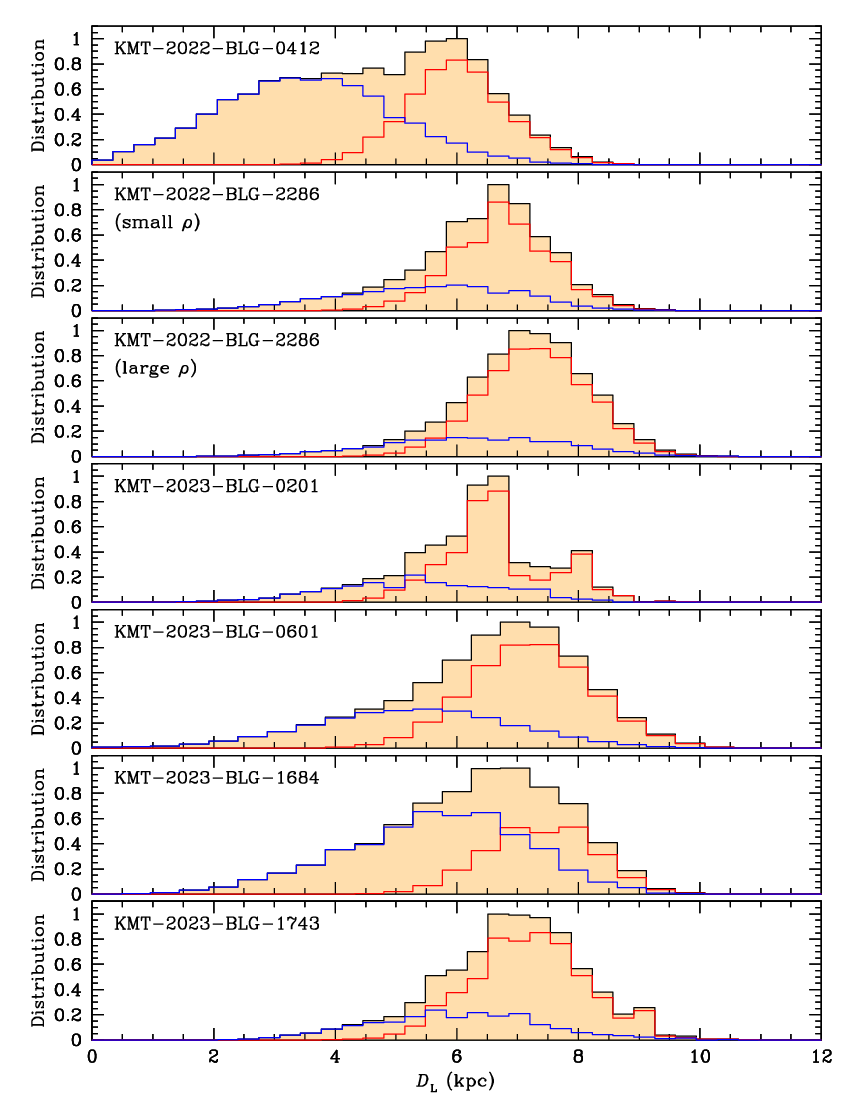}
\caption{
Bayesian posteriors for the distance to the lens system. The notations are the same as
those in Fig.~\ref{fig:fifteen}.
}
\label{fig:sixteen}
\end{figure}

\begin{table*}[t]
\caption{Physical lens parameters.\label{table:eleven}}
\begin{tabular}{lllllccll}
\hline\hline
\multicolumn{1}{c}{Event}                  &
\multicolumn{1}{c}{$M_1$ ($M_\odot$) }     &
\multicolumn{1}{c}{$M_2$ ($M_\odot$) }     &
\multicolumn{1}{c}{$\dl$ (kpc) }           &
\multicolumn{1}{c}{$a_\perp$ (au) }        &
\multicolumn{1}{c}{$p_{\rm disk}$ (\%)}    &
\multicolumn{1}{c}{$p_{\rm bulge}$ (\%)}  \\
\hline
 KMT-2022-BLG-0412                &  $0.65^{+0.38}_{-0.37}$   &  $0.040^{+0.023}_{-0.023}$    &   $4.82^{+1.63}_{-2.10}$  &   $2.86^{+0.97}_{-1.25}$   &   61   &   39  \\  [0.5ex]
 KMT-2022-BLG-2286 (small $\rho$) &  $0.45^{+0.33}_{-0.24}$   &  $0.041^{+0.029}_{-0.022}$    &   $6.71^{+0.91}_{-1.33}$  &   $3.32^{+0.45}_{-0.66}$   &   33   &   67  \\  [0.5ex]
 \hskip95pt (large $\rho$)        &  $0.20^{+0.29}_{-0.11}$   &  $0.018^{+0.026}_{-0.010}$    &   $7.32^{+0.97}_{-1.12}$  &   $2.02^{+0.27}_{-0.31}$   &   23   &   77  \\  [0.5ex]
 KMT-2023-BLG-0201                &  $0.26^{+0.30}_{-0.14}$   &  $0.052^{+0.061}_{-0.028}$    &   $6.55^{+1.13}_{-1.24}$  &   $2.05^{+0.35}_{-0.39}$   &   32   &   68  \\  [0.5ex]
 KMT-2023-BLG-0601                &  $0.51^{+0.40}_{-0.33}$   &  $0.033^{+0.026}_{-0.021}$    &   $6.95^{+1.27}_{-1.87}$  &   $1.56^{+0.29}_{-0.42}$   &   38   &   62  \\  [0.5ex]
 KMT-2023-BLG-1684                &  $0.68^{+0.34}_{-0.36}$   &  $0.046^{+0.023}_{-0.024}$    &   $6.66^{+1.36}_{-1.80}$  &   $2.48^{+0.50}_{-0.67}$   &   65   &   35  \\  [0.5ex]
 KMT-2023-BLG-1743                &  $0.11^{+0.16}_{-0.06}$   &  $0.019^{+0.029}_{-0.010}$    &   $7.13^{+1.05}_{-1.27}$  &   $1.16^{+0.17}_{-0.21}$   &   27   &   73  \\  [0.5ex]
\hline                                                                                                       
\end{tabular}
\end{table*}

\section{Physical lens parameters}\label{sec:six}

The physical parameters of a lens are constrained by the lensing observables $\te$ and $\thetae$. 
These observables are related to the lens parameters of the mass $M$ and distance $\dl$ by
\begin{equation}
\te = {\thetae \over \mu};\qquad     \thetae = \sqrt{\kappa M \pi_{\rm rel}},
\label{eq4}
\end{equation}
where $\kappa  = 4G/(c^2{\rm au})$, 
$\pi_{\rm rel}={\rm au}(1/\dl-1/\ds)$ is the relative lens-source parallax, and $\ds$ represents 
the distance to the source.  In a special case where an additional observable of the microlens parallax 
$\pie$ is measured, the lens mass and distance are uniquely determined by the relations;
\begin{equation}
M = {\thetae \over \kappa \pie};\qquad  
\dl = {{\rm au} \over \pie\thetae + \pi_{\rm S}}.
\label{eq5}
\end{equation}
Here $\pi_{\rm S} = {\rm au}/\ds$ represents the parallax of the source.  The observables for the 
analyzed events were partially measured: the event timescale was determined for all events, but 
the angular Einstein radius was not measured for KMT-2022-BLG-0412 and KMT-2023-BLG-0601, and the 
microlens parallax could not be reliably determined for any of the events.  Therefore, we estimate 
the physical lens parameters by conducting a Bayesian analysis using a prior Galaxy model along with 
the constraints provided by the measured observables of the events.

The Bayesian analysis was conducted as follows. Initially, we generated a large number of synthetic
events through a Monte Carlo simulation. In this process, we determined the lens mass ($M_i$) from
a model mass function and inferred the distances to the lens and source $(D_{{\rm L},i}, D_{{\rm S},i})$
as well as their relative proper motion ($\mu_u$), using a Galaxy model. We employed the mass function 
model proposed by \citet{Jung2018} and the Galaxy model introduced by \citet{Jung2021}. For each simulated 
event characterized by a set of the physical parameters $(M_i, D_{{\rm L},i}, D_{{\rm S},i}, \mu_i)$, we 
calculated the lensing observables of the timescale and angular Einstein radius by applying the equations 
in Eq.~(\ref{eq4}). In the subsequent step, we constructed the posterior distribution of the lens mass and 
distance by assigning a weight to each simulated event of
\begin{equation}
w_i = \exp\left(-{\chi_i^2 \over 2} \right); \qquad
\chi_i^2 = { (t_{{\rm E},i}-\te)^2\over \sigma(\te)^2} + { (\theta_{{\rm E},i}-\theta)^2\over \sigma(\theta)^2}.
\label{eq6}
\end{equation}
Here $(\te, \thetae)$ denote the measured values of the lensing observables and $[\sigma(\te), \sigma(\thetae)]$ 
represent their uncertainties.

Figures~\ref{fig:fifteen} and \ref{fig:sixteen} display the Bayesian posteriors constructed for the companion 
lens mass and distance. For event KMT-2022-BLG-2286, which has two degenerate solutions with different $\thetae$ 
values, we provide two sets of posteriors. Table~\ref{table:ten} summarizes the estimated masses of the primary 
lens ($M_1$) and its companion ($M_2$), along with the distance and the projected separation between $M_1$ and 
$M_2$ ($a_\perp$). The projected separation is computed as $a_\perp=s\dl\thetae$. The representative value is 
presented as the median of the posterior distribution, with the lower and upper limits estimated at the 16th 
and 84th percentiles, respectively.

An examination of the mass posteriors shows that the median masses of the lens companions for all events fall 
within the brown dwarf mass range of $13M_{\rm J} \leq M_2 \leq 80M_{\rm J}$.  In terms of median posterior 
values, the mass of the primary lens ranges from $0.11~M_\odot$ to $0.65~M_\odot$, indicating that they are 
low-mass stars with substantially lower masses compared to the Sun. The table also provides the probabilities 
for the lens being in the disk, $p_{\rm disk}$, and in the bulge, $p_{\rm bulge}$.  Specifically, for event 
KMT-2022-BLG-0412, the lens is more likely to be located in the disk. In contrast, for events KMT-2022-BLG-2286, 
KMT-2023-BLG-0201, KMT-2023-BLG-0601, and KMT-2023-BLG-1743, the lenses are more likely to be situated in the 
bulge. For event KMT-2023-BLG-1684, the probabilities of the lens being in the disk or the bulge are similar.

\begin{table*}[t]
\footnotesize
\caption{Relative proper motion, angular separation in 2030, and $K$-band source magnitude.\label{table:twelve}}
\begin{tabular}{lllll}
\hline\hline
\multicolumn{2}{c}{Event}                         &
\multicolumn{1}{c}{$\mu$ (mas/yr)}                &
\multicolumn{1}{c}{$\Delta\theta_{2030}$ (mas)}   &
\multicolumn{1}{c}{$K$ (mag)}                     \\
\hline 
\citet{Gould2022}     & OGLE-2016-BLG-0596               &  $5.1 \pm 0.8 $    &  $71.4 $   &  $17.89$  \\
                      &  OGLE-2016-BLG-0693              &  $1.5 \pm 0.2 $    &  $21.0 $   &  $18.64$  \\
                      &  OGLE-2016-BLG-1190              &  $1.9 \pm 0.2 $    &  $26.6 $   &  $18.69$  \\
                      &  KMT-2016-BLG-0212               &  $8.1 \pm 2.5 $    &  $113.4$   &  $18.41$  \\
                      &  KMT-2016-BLG-1107               &  $2.6 \pm 0.4 $    &  $36.4 $   &  $12.79$  \\
                      &  KMT-2016-BLG-2605               &  $12.3 \pm 1.0$    &  $172.2$   &  $16.58$  \\
                      &  OGLE-2017-BLG-1375              &  $3.6 \pm 0.5 $    &  $46.8 $   &  $19.28$  \\
                      &  OGLE-2017-BLG-1522              &  $3.2 \pm 0.5 $    &  $41.6 $   &  $19.34$  \\
                      &  OGLE-2018-BLG-1011              &  $2.8 \pm 0.2 $    &  $33.6 $   &  $16.65$  \\
                      &  OGLE-2018-BLG-1647              &  $0.6 \pm 0.1 $    &  $7.2  $   &  $18.51$  \\
                      &  OGLE-2019-BLG-0468              &  $4.4 \pm 0.6 $    &  $48.4 $   &  $18.73$  \\
                      &  KMT-2019-BLG-0371               &  $7.7 \pm 0.7 $    &  $84.7 $   &  $17.17$  \\
\hline
Paper I               &  OGLE-2016-BLG-0890              &  $6.30 \pm 1.12 $  &  $88.2  $  &  $12.44 $  \\
                      &  MOA-2017-BLG-477                &  $9.33 \pm 0.83 $  &  $121.29$  &  $18.21 $  \\
                      &  KMT-2018-BLG-0357               &  $7.52 \pm 1.05 $  &  $90.24 $  &  $17.81 $  \\
                      &  OGLE-2018-BLG-1489              &  $4.89 \pm 0.36 $  &  $58.68 $  &  $16.50 $  \\
                      &  OGLE-2018-BLG-0360              &  $4.12 \pm 0.59 $  &  $49.44 $  &  $16.83 $  \\
\hline                                                                                   
Paper II              &  KMT-2018-BLG-0321               &  $> 2.4         $  &  $> 28.8$  &  $16.05 $  \\
                      &  KMT-2018-BLG-0885               &  $> 1.6         $  &  $> 19.2$  &  $18.58 $  \\
                      &  KMT-2019-BLG-0297               &  $6.93 \pm 0.51 $  &  $76.23 $  &  $17.88 $  \\
                      &  KMT-2019-BLG-0335               &  $2.87 \pm 1.16 $  &  $31.57 $  &  $18.89 $  \\
\hline                                                                                    
Paper III             &  KMT-2021-BLG-0588               &  $8.68 \pm 0.80 $  &  $78.12 $  &  $17.68 $  \\
                      &  KMT-2021-BLG-1110               &  $7.08 \pm 0.62 $  &  $63.72 $  &  $18.75 $  \\
                      &  KMT-2021-BLG-1643               &  $3.76 \pm 0.58 $  &  $33.84 $  &  $19.51 $  \\
                      &  KMT-2021-BLG-1770               &  $7.63 \pm 0.90 $  &  $68.67 $  &  $17.08 $  \\
\hline                
This work             &  KMT-2022-BLG-0412               &  $ > 1.1        $  &  $ > 8.8$  &  $18.12$  \\
                      &  KMT-2022-BLG-2286 (small $\rho$)&  $8.97 \pm 0.14 $  &  $71.7  $  &  $16.93$  \\
                      &  \hskip84pt   (large $\rho$)     &  $4.59 \pm 0.56 $  &  $36.7  $  &  $16.93$  \\
                      &  KMT-2023-BLG-0201               &  $10.30 \pm 0.97$  &  $72.1  $  &  $12.97$  \\
		      &  KMT-2023-BLG-1684               &  $2.82 \pm 0.51$   &  $19.7  $  &  $13.31$  \\
	              &  KMT-2023-BLG-1743               &  $9.60 \pm 0.79 $  &  $67.2  $  &  $15.40$  \\
\hline
\end{tabular}
\end{table*}

\section{Summary and discussion}\label{sec:seven}

Building on the groundwork laid in Papers I through III, which aimed to construct a homogeneous sample 
of brown dwarfs in binary systems, we extended our investigation to microlensing events detected by 
the KMTNet survey during the 2022 and 2023 seasons.  Due to the challenge of distinguishing brown 
dwarf events from those produced by binary lenses with nearly equal-mass components, we analyzed all 
lensing events detected during these seasons that exhibited anomalies characteristic of binary-lens systems.  
By applying the same criteria consistently used in previous studies, we identified seven additional brown 
dwarf candidates through the analysis of the following lensing events: KMT-2022-BLG-0412, KMT-2022-BLG-2286, 
KMT-2023-BLG-0201, KMT-2023-BLG-0601, KMT-2023-BLG-1684, and KMT-2023-BLG-1743.

Based on the Bayesian analysis performed with the constraints derived from measured lensing observables, 
it was determined that the median mass of the lens companions spans from $0.02~M_\odot$ to $0.05~M_\odot$, 
affirming their placement within the brown dwarf mass spectrum. Additionally, the mass range of the primary 
lenses extends from $0.11~M_\odot$ to $0.65~M_\odot$, suggesting that they are low-mass stars with 
significantly lesser masses compared to the Sun.

The BD nature of the lens companions can be verified through direct lens imaging from future high-resolution 
adaptive optics (AO) follow-up observations. This verification will be feasible once the lenses are 
sufficiently separated from the source stars. Around 2030, the AO observations with the 30-meter European 
Extremely Large Telescope are expected to resolve the lenses from the source stars.  This will enable to 
estimate the mass of the primary lens. Combined with the accurately measured mass ratio, the BD nature 
of the lens companion can be confirmed.

\begin{figure}[t]
\includegraphics[width=\columnwidth]{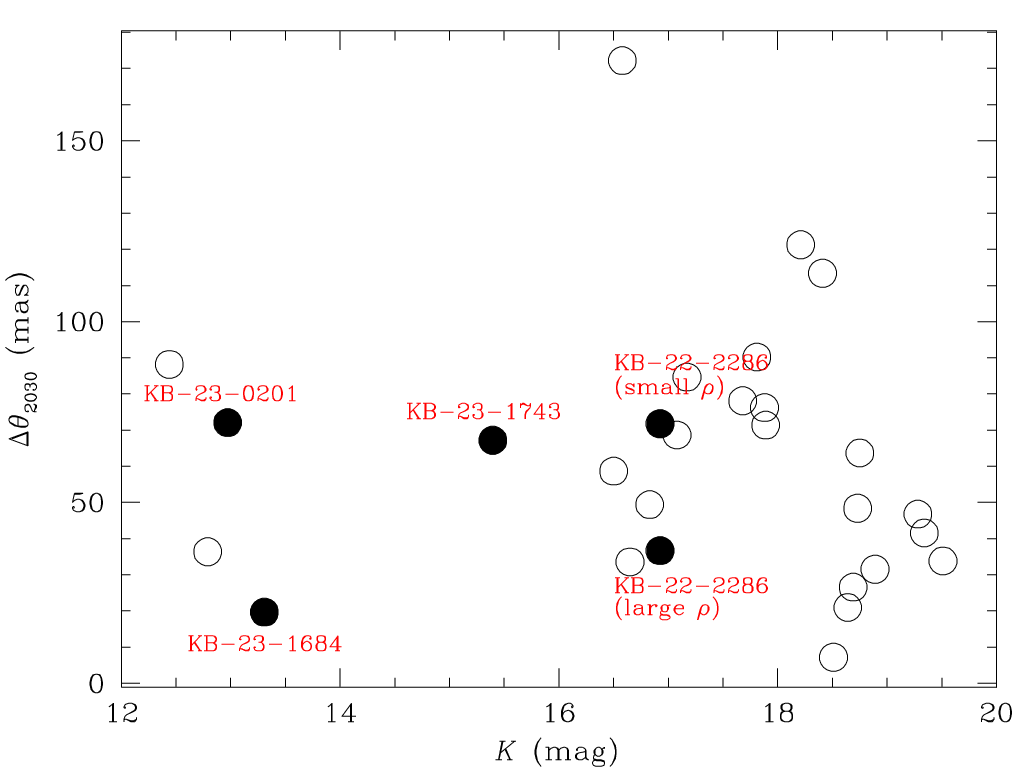}
\caption{
BD candidate events in the the plane of the $K$-band magnitude and $\Delta\theta_{2030}$.
The events analyzed in this work are marked by filled dots.
}
\label{fig:seventeen}
\end{figure}

Table~\ref{table:twelve} contains the necessary information for preparing these follow-up observations. 
This includes the expected separations between the lens and source in 2030 ($\Delta\theta_{2030}$), the 
$K$-band source magnitudes, and the values of the relative lens-source proper motion. The table covers 
events discussed in this work as well as those published in Papers I through III.  Additionally, \citet{Gould2022} 
provided similar tables for over 100 planetary lensing events. Of these, 12 events have companion/primary mass 
ratios that meet our BD selection criterion ($10^{-2} < q < 10^{-1}$). Information on these events is included 
as well.  For KMT-2023-BLG-0601, information on $\Delta\theta_{2030}$ is not presented because the value of 
the relative proper motion was not measured.  In Figure~\ref{fig:seventeen}, we plot the locations of the 
events in the plane of the $K$-band magnitude and $\Delta\theta_{2030}$.  According to the table, the 
lens-source separation will exceed 50 mas for KMT-2023-BLG-0201 and KMT-2023-BLG-1743, indicating that the 
lenses can be readily resolved with AO observations.  In contrast, the expected separation for KMT-2023-BLG-1684, 
$\Delta\theta_{2030}\sim 20$~mas, is substantially smaller, suggesting difficulty in resolving the lens.  For 
KMT-2022-BLG-2286, the expected separations for the small-$\rho$ solution ($\Delta\theta_{2030}\sim 72$~mas) 
and the large-$\rho$ solution ($\Delta\theta_{2030}\sim 37$~mas) differ significantly, indicating that AO 
follow-up will not only easily resolve the degeneracy but also yield a host-mass measurement.

\begin{acknowledgements}
Work by C.H. was supported by the grants of National Research Foundation of Korea (2019R1A2C2085965). 
J.C.Y. and I.-G.S. acknowledge support from U.S. NSF Grant No. AST-2108414. 
Y.S. acknowledges support from BSF Grant No. 2020740.
This research has made use of the KMTNet system operated by the Korea Astronomy and Space
Science Institute (KASI) at three host sites of CTIO in Chile, SAAO in South Africa, and SSO in
Australia. Data transfer from the host site to KASI was supported by the Korea Research
Environment Open NETwork (KREONET). This research was supported by KASI under the R\&D
program (Project No. 2024-1-832-01) supervised by the Ministry of Science and ICT.
W.Z. and H.Y. acknowledge support by the National Natural Science Foundation of China (Grant
No. 12133005).
W. Zang acknowledges the support from the Harvard-Smithsonian Center for Astrophysics through
the CfA Fellowship. 
The MOA project is supported by JSPS KAKENHI Grant Number JP24253004, JP26247023,JP16H06287 and JP22H00153.
\end{acknowledgements}

\end{document}